\documentclass[journal]{IEEEtran}
\usepackage{amsmath,amsfonts}
\usepackage{algorithmic}
\usepackage{algorithm}
\usepackage{array}
\usepackage[caption=false,font=normalsize,labelfont=sf,textfont=sf]{subfig}
\usepackage{textcomp}
\usepackage{stfloats}
\usepackage{url}
\usepackage{verbatim}
\usepackage{graphicx}
\usepackage{cite}
\usepackage{paralist}
\usepackage{dsfont}
\usepackage{amsthm} 
\usepackage{amssymb} 
\usepackage{bm}      
\usepackage{multicol}
\usepackage{hyperref}
\usepackage{xcolor}
\usepackage{xr} %
\usepackage{booktabs}
\usepackage{caption}
\usepackage{color,array,amsthm}
\usepackage{graphicx}

\newtheorem{proposition}{Proposition}

\begin{document}

\title{Joint Inverse Learning of Cognitive Radar Perception and Perception-Action Policy}

\author{
    Anoop C V and
    \thanks{Both Anoop C V and Anup Aprem are with the Department of Electronics and Communication Engineering, National Institute of Technology Calicut, India (e-mail: \url{anoop_p210065ec@nitc.ac.in}, \url{anup.aprem@nitc.ac.in}).}
    Anup Aprem
}

\maketitle

\begin{abstract}Cognitive Radars (CRs) employ perception–action cycle to adapt their sensing and transmission strategies based on its' perception of the target kinematic states and mission objectives. 
This paper considers an inverse learning Electronic Counter Measure (ECM) that infers both the perception and perception-driven action policy of the adversarial CR’s from the actions of the CR, i.e. the sensing and transmission actions taken by the CR.    
Existing frameworks, in the literature, assume the knowledge of either the perception or the perception-action policy and infer the other. However, this assumption is unrealistic in an adversarial setting.  
We address this gap by proposing an online, nonparametric Bayesian machine learning framework and developing the Inverse Particle Filter with Dependent Dirichlet Process (IPFDDP) algorithm, which characterizes the perception-dependent action policy using a Dependent Dirichlet Process (DDP) and embeds kernel-based DDP inference within a Bayesian inverse particle filtering framework to jointly estimate the CR’s perception and perception-action policy. Extensive numerical simulations demonstrate that IPFDDP outperforms existing inverse learning methods in terms of mean squared error, Kullback–Leibler divergence between the estimated and true policy, and accuracy in identifying relative action preferences.  
Unlike the existing techniques, the proposed Bayesian formulation naturally quantifies uncertainty in inferred perception and perception-action policy, enabling active probing strategies for sample efficient inverse learning. Simulation results show that active probing integrated with IPFDDP achieves, on average, a $40\%$ faster reduction in KL divergence compared to randomized probing.
\end{abstract}


\section{Introduction}

\IEEEPARstart{C}{ognitive} Radars (CRs) are intelligent sensing systems capable of environmental awareness and autonomous adaptation, allowing them to adjust their sensor parameters or policies adaptively for more accurate target tracking and detection.\cite{haykin2006cognitive}. They adjust various operational parameters, such as waveform characteristics, beam direction, transmission power, and track update periodicity, based on their perception of the target state, aligned with its' operational objectives or mission intent specified by the operators. These mission objectives frequently include, but are not limited to, minimizing tracking uncertainty, lowering the false alarm rate in target detection, ensuring a low probability of interception (LPI), and facilitating coexistence with other co-located communication systems \cite{bockmair2019cognitiveIEEEAE}.

Electronic Counter Measure (ECM) denotes a collection of offensive and/or defensive strategies to either deny an \emph{adversarial} radar accurate targeting information or exploit vulnerabilities in the capabilities of radar. 
This paper considers an electronic warfare context wherein an intelligent surveillance agent tries to infer the radar's perception and the perception driven policy of the CR - referred to as \emph{inverse learning}\cite{ECM2023radar,Vikram2022ECCM} with an intent to design ECM strategies to exploit vulnerabilities in the perception-action policy of the CR. 
This inverse learning challenge has received considerable attention in the literature, particularly in the context of CRs\cite{krishnamurthy2019calibrate,IPF2025IITD,KM_Patta2021CRmodel,AnoopHBOIL}.

A CR employs Bayesian filtering techniques (e.g., Kalman Filter, Extended Kalman Filter, Unscented Kalman Filter, or Particle Filter) to obtain the posterior belief, which represents the radar's perception of the target's kinematic state, conditional on noisy observations \cite{MatFrame5}. Subsequently, it executes adaptive actions (such as modifying waveform parameters or adjusting beam steering) as a function of this posterior belief \cite{sira2022advances,CODIR,krishnamurthy2020CR_RP}. For example, the study in \cite{sira2007dynamic,sira2022advances} demonstrates that the FAR dynamically selects one of three available waveforms—Linear Frequency Modulation (LFM), Power Frequency Modulation (PFM), or Hyperbolic Frequency Modulation (HFM)—with the objective of minimizing the tracking covariance produced by its internal Bayesian tracker. 

In the context of inverse learning in the CR, Bayesian inverse filtering approaches in the literature \cite{krishnamurthy2019calibrate,IPF2025IITD} estimate the radar's perception of the target (estimate of the posterior belief), assuming that the perception adaptation policy of the radar is known apriori.
The estimated perception can then be utilized to predict the radar's subsequent actions under a novel set of environmental conditions. 
The assumption of perfect knowledge of the perception driven action policy of an adversarial CR is unrealistic in an electronic warfare scenario or in dynamic environments.

\begin{figure*}
    \centering
    \includegraphics[width=\linewidth]{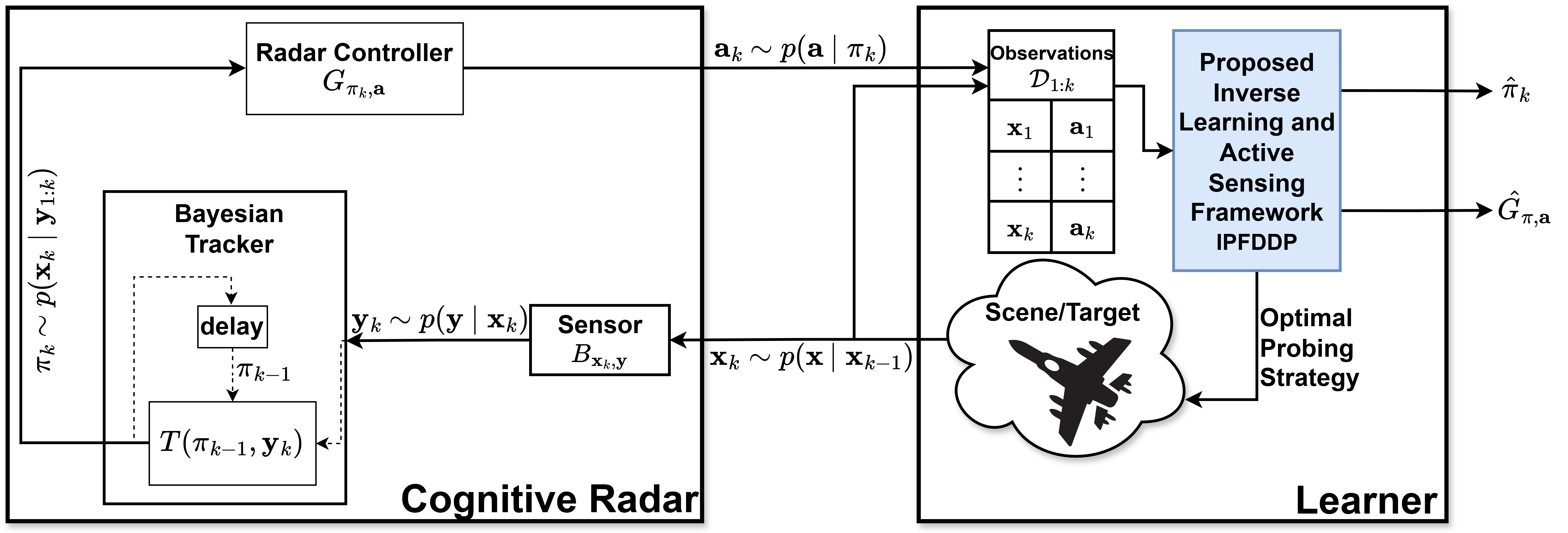}
    \caption{CR-target interaction and the proposed inverse learning approach. We consider the problem of jointly estimating the radar's belief state \(\pi_k\) and its belief-dependent policy \(G_{\pi,\mathbf{a}} := p(\mathbf{a}|\pi)\), solely from observed state trajectories \(\mathbf{x}_{1:k}\) and corresponding radar actions \(\mathbf{a}_{1:k}\). The main challenge arises from the fact that the radar's belief \(\pi_k\) is itself unobservable, and the action policy \(G_{\pi,\mathbf{a}}\) is also unknown and potentially nonparametric.}\label{fig:IL_BD}
\end{figure*}

In this paper, we address the main limitation of assumption of perception-action policy in existing literature. The main challenge arises from the fact that the radar's perception (belief) of the target state is itself unobservable.
Subsequently inverse learner must jointly estimate the radar's perception and its perception-dependent policy, from observed state trajectories and corresponding radar actions.
In this context, the main contributions of this paper are summarized as follows:
\begin{enumerate}
    \item We propose an online, nonparametric Bayesian machine learning framework for jointly inferring the perception and the perception-dependent stochastic action policy of a cognitive radar, by observing a sequence of actions taken by the CR corresponding to known kinematic states of the target.
    \item We develop IPFDDP, a nonparametric Bayesian machine learning algorithm to jointly infer the policy (or belief) and perception-action policy of the CR. The IPFDDP algorithm characterizes the belief-dependent policy of the CR using the Dependent Dirichlet Process (DDP), and incorporates kernel-based DDP inference into a Bayesian inverse particle filtering framework to jointly estimate the CR's belief and policy.
    \item We compare the performance of the IPFDDP-based policy learning algorithm with state-of-the-art ML models, based on MSE and KL divergence between the predicted and the true policy, as well as accuracy in identifying the relative preferences for actions, and show that IPFDDP outperforms the competing approaches across all metrics.
    \item The proposed Bayesian formulation inherently quantifies the uncertainty in the learner's estimates of belief and policy and facilitates the design of active probing strategies to perform inference in a sample-efficient manner. We propose an active sensing approach in which the target on the learner's side adapts its acceleration maneuver to probe the CR to minimize uncertainty in the inverse learning estimate.
    \item Numeric simulations show that incorporating active probing strategy into IPFDDP reduces the KL divergence between the estimated and true policy on an average by \(40\%\) faster compared to randomized probing.
\end{enumerate}
\emph{To the best of our knowledge, there exist no learning frameworks in the literature that can jointly estimate both the perception and the perception-action policy of a cognitive radar.}

This paper is organized as follows. 
Sec.~\ref{sec:LitRev} reviews the related literature on inverse learning based ECM strategies. 
Sec.~\ref{sec:Background} presents the system model and outlines the necessary background, including Bayesian inverse filtering and the Dirichlet process formulation in the considered context. 
Sec.~\ref{sec:PropMeth} details the proposed inverse learning framework and the associated active sensing strategies. 
Comprehensive numerical results are provided in Sec.~\ref{sec:Results}. 
Further, conclusions are drawn in Section~\ref{sec:Conclusion}. 
The detailed architectures and training procedures of the machine learning models employed for benchmarking in Sec.~\ref{sec:Results} are provided as supplementary material.

\begin{table*}[t]
\centering
\caption{Knowledge, Observability and Assumptions of inverse learner in the literature}
\begin{tabular}{lllll}
\toprule
\textbf{Quantity} & \textbf{Description} & \textbf{IF\cite{krishnamurthy2019calibrate,IEKF_2023_Chattopad,IUKF_2024_Chattopad,IPF2025IITD}} & \textbf{IRL\cite{krishnamurthy2020CR_RP,AnoopHBOIL}} & \textbf{Proposed} \\
\midrule
\(P, B\)             & Model dynamics          & Known        & Known        & Known \\
\(\mathbf{x}_{0:k}\) & Target state trajectory & Observable   & Observable   & Observable \\
\(\mathbf{a}_{1:k}\) & Radar action trajectory & Observable   & Observable   & Observable \\
\(\mathbf{y}_{1:k}\) & Radar's measurements    & Unobservable & Unobservable & Unobservable \\
\hline
\(\pi_{1:k}\)        & Radar belief state      & \textbf{Unobservable} & Known & \textbf{Unobservable} \\
\(G_{\pi,\mathbf{a}}\) & Belief-dependent radar policy & Known & \textbf{Unknown} & \textbf{Unknown} \\
\(\mathcal{A}\)      & Action set     & Known        & Known        & \textbf{Unknown} \\
\bottomrule
\end{tabular}
\label{tab:assumptions}
\end{table*}
\section{Related Literature}\label{sec:LitRev}
{\sl \textbf{Classical ECM}: }ECM against conventional radars include a set of pre-programmed techniques such as noise jamming, deceptive jamming (false target generation) and repeated based spoofing \cite{neri2006EWintroduction,skolnik2008radar}, which  assume stationary radar operating modes and are therefore not suitable for cognitive radars that continuously adapt waveform, beamforming, and processing based on feedback from the scene (the environment in which the target is present) \cite{haigh2021cognitiveEW_AI}. %
ECM against cognitive radar must therefore adaptively adjust its strategy based on the behaviour of the cognitive radar learned from observations \cite{haykin2006cognitive,haigh2021cognitiveEW_AI}. %

{\sl \textbf{Bayesian Inverse Filtering (IF) based ECM}: }Inverse filtering techniques to aid ECM attempt to learn the radar's posterior state estimate or perception of the scene (e.g., posterior of the target state estimate or track quality) by inverting its Bayesian filtering structure. \cite{krishnamurthy2019calibrate} introduced Inverse Hidden Markov Model Filter (IHMMF), Inverse Kalman Filter (IKF) and Inverse Particle Filter (IPF) in the context of ECM against CRs. Inverse Extended Kalaman Filter (IEKF) \cite{IITBombayIEKF,IEKF_2023_Chattopad}, Inverse Unscented Kalman Filters (IUKF) \cite{IUKF_2024_Chattopad} and Inverse Particle Filter (IPF) extends the work in \cite{krishnamurthy2019calibrate} to a non-linear setting. But, all the existing Bayesian Inverse Filtering works in literature \cite{krishnamurthy2019calibrate,IITBombayIEKF,IEKF_2023_Chattopad,IUKF_2024_Chattopad,IPF2025IITD} typically assume that the radar's action policy is known apriori---unrealistic in an electronic warfare scenario or in dynamic environments. %

{\sl \textbf{Inverse Reinforcement Learning (IRL) based ECM: }}In contrast, inverse learning / inverse reinforcement learning based approaches \cite{krishnamurthy2020CR_RP,AnoopHBOIL} in the literature assume that the radar perceives the state of the target perfectly and that the actions taken by the radar are optimal with respect to some reward / utility function, and subsequently attempt to estimate the reward / utility that the radar seeks to maximize. Similarly, inverse reinforcement learning, inverse optimal control, and changepoint-detection-based techniques proposed to aid ECM also assume that the radar perceives the target state perfectly, or is provided with ground-truth state trajectories, and focus on recovering the radar's reward structure, utility, or mode-switching strategy from observed actions \cite{krishnamurthy2020cognitive,AnoopHBOIL}. However, an adversarial radar's perception of the target state is generally not known \textit{perfectly} to the learner and must therefore be inferred. Moreover, the inverse learning performance of parametric approaches is sensitive to the assumed functional form of the reward / utility\cite{jarrett2021IDM}. 

In addition, naive neural network based approximations of the mapping from the true kinematic states of the targets to radar actions typically fail or are not sample efficient, since the radar's actions are not a direct function of the true target states but rather of the CR's perception of the target states, which is unknown to the learner, and because the number of observations available for inference in adversarial settings is often scarce \cite{AnoopHBOIL}. Furthermore, modern CRs with meta-cognitive capabilities often take suboptimal or stochastic decisions either to mask their cognition from adversaries \cite{krishnamurthy2023meta} or to enable better exploration under imperfect perception of the environment. Consequently, the assumptions underlying both inverse filtering and inverse reinforcement learning approaches are not valid in realistic radar--inverse learner engagements, where the radar operates under partial observability and model uncertainty, and both the radar's perception (state estimation) and policy (waveform / processing selection) must be inferred jointly from observations of its actions, as summarized in Table~\ref{tab:assumptions}.

\section{Background and System Model}\label{sec:Background}

We consider an adversarial interaction between a CR and a learner, as illustrated in Fig.~\ref{fig:IL_BD}. The CR tracks a maneuvering target (drone/UAV/aircraft) located on the learner’s side. The learner observes the radar’s actions and seeks to infer its internal perception and perception-action policy. The overall architecture consists of a scene (target dynamics), a sensor, a Bayesian tracker, and a perception-driven radar controller.

{\sl \textbf{Scene and Target Dynamics}: }
The scene consists of a moving target whose kinematic state evolves in discrete time $k=1,2,\ldots$. Let $\mathbf{x}_k \in \mathcal{X}$ denote the Markovian state of the target at time $k$, with prior $\mathbf{x}_0 \sim \pi_0$. The motion dynamics are governed by
\begin{align}
\textbf{Motion dynamics:}\;
& \mathbf{x}_k \sim P_{\mathbf{x}_{k-1},\mathbf{x}_k}
  = p(\mathbf{x}_k \mid \mathbf{x}_{k-1}),\label{eq:dynamics_mot}
\end{align}
where $\mathcal{X}$ denotes the state space and $P_{\mathbf{x}_{k-1},\mathbf{x}_k}$ specifies the target motion model.

{\sl \textbf{Sensor Model}: }
The CR observes the target state through noisy measurements $\mathbf{y}_k$ generated according to
\begin{align}
\textbf{Observation:}\;
& \mathbf{y}_k \sim B_{\mathbf{x}_k,\mathbf{y}}
  = p(\mathbf{y}_k \mid \mathbf{x}_k),\label{eq:dynamics_obs}
\end{align}
where $B_{\mathbf{x}_k,\mathbf{y}}$ denotes the conditional probability density of the observation given the true target state.

{\sl \textbf{Bayesian Tracker}: }
The CR employs a Bayesian tracker (e.g., Kalman filter or particle filter) to compute a posterior belief over the target state. Let $\pi_k \in \Pi$ denote the posterior belief at time $k$. The belief evolves according to the Bayesian filtering recursion
\begin{align}
\textbf{Perception:}\;
& \pi_k = T(\pi_{k-1},\mathbf{y}_k)
  = p(\mathbf{x}_k \mid \mathbf{y}_{1:k}),\label{eq:dynamics_bel}
\end{align}
where $\mathbf{y}_{1:k} = [\mathbf{y}_1,\ldots,\mathbf{y}_k]$ and the filtering operator $T$ is given by
\begin{equation}\label{eq:BayesFilter}
    T(\pi, \mathbf{y}) = \frac{B_{\mathbf{x},\mathbf{y}}\int_{\mathcal{X}}P_{\zeta,\mathbf{x}}\pi(\zeta)d\zeta}{\int_{\mathcal{X}}B_{\mathbf{x},\mathbf{y}}\int_{\mathcal{X}}P_{\zeta,\mathbf{x}}\pi(\zeta)d\zeta d\mathbf{x}}, \; \mathbf{x} \in \mathcal{X}.
\end{equation}

Thus, $\pi_k$ represents the internal perception of the CR regarding the target state.

{\sl \textbf{Radar Controller}}
Based on the belief $\pi_k$, the radar controller selects an action $\mathbf{a}_k \in \mathcal{A}$ according to a perception-dependent stochastic policy
\begin{align}
\hspace{-0.7em}\textbf{Perception--action policy:}\;
& \mathbf{a}_k \sim G_{\pi_k,\mathbf{a}}
  = p(\mathbf{a}_k \mid \pi_k).\label{eq:dynamics_act}
\end{align}
Here, $\mathcal{A}$ denotes the radar action space (e.g., waveform selection, beam steering, or power allocation), and $G_{\pi_k,\mathbf{a}}$ characterizes the conditional distribution of actions given the belief. We assume that the belief space $\Pi$ is partitioned into disjoint regions, with distinct policies associated with each region.

In this work, the learner is assumed to perfectly observe the radar actions. This assumption is reasonable when actions correspond to discrete waveform switching. Extensions to scenarios where actions are observed through noise (e.g., power adaptation) are deferred to future work.

{\sl \textbf{Information Structure and Inverse Learning Problem: }}
As shown in Fig.~\ref{fig:IL_BD}, the CR internally implements the forward generative chain
\[
 \mathbf{x}_k \rightarrow \mathbf{y}_k \rightarrow \pi_k \rightarrow \mathbf{a}_k,
\]
where the belief $\pi_k$ is a sufficient statistic of the observation history for decision-making.

The learner does not observe the radar measurements $\mathbf{y}_k$ nor the belief trajectory $\pi_{0:k} = [\pi_0,\pi_1, \cdots, \pi_k]$. Instead, it observes the target state trajectory $\mathbf{x}_{0:k}= [\mathbf{x}_0,\mathbf{x}_1, \cdots, \mathbf{x}_k]$ on its side and the radar action trajectory $\mathbf{a}_{1:k}= [\mathbf{a}_1,\mathbf{a}_2,\cdots,\mathbf{a}_k]$. 

The objective of the learner is to solve the following inverse problem: given the observed state-action sequence $(\mathbf{x}_{0:k}, \mathbf{a}_{1:k})$, jointly infer the hidden belief process $\pi_{0:k}$ and the perception-dependent stochastic policy $G_{\pi,\mathbf{a}}$. This constitutes an adversarial inverse learning problem in which the learner attempts to reconstruct both the internal perception dynamics and the control strategy of the enemy CR.

In Sec.~\ref{sec:Bayesian_IF}, we discuss the classical Bayesian inverse filtering, where the learner recursively estimates the CR's posterior belief of its kinematic state, given \(\mathbf{a}_{1:k}\) and \(\mathbf{x}_{0:k}\), under the assumptions that \(P, B, \mathcal{A}\), and \(G_{\pi,\mathbf{a}}\) in \eqref{eq:dynamics_mot} through \eqref{eq:dynamics_act} are perfectly known to the learner, and in Sec.~\ref{sec:PropMeth}, we extend to the case of unknown \(\mathcal{A}\) and \(G_{\pi,\mathbf{a}}\).

\subsection{Bayesian Inverse Filtering}\label{sec:Bayesian_IF} 
Based on the model dynamics in \eqref{eq:dynamics_mot} through \eqref{eq:dynamics_act}, the joint distribution of \(\pi_k, \mathbf{y}_k,\mathbf{a}_{1:k}\) and $\mathbf{x}_{0:k}$ can be expressed as:
\begin{align}
    &p(\pi_k, \mathbf{y}_k,\mathbf{a}_{1:k}, \mathbf{x}_{0:k}) = p(\mathbf{a}_k\mid \pi_k)p(\mathbf{y}_k\mid \mathbf{x}_k)p(\mathbf{x}_k\mid\mathbf{x}_{k-1})\notag\\&
    \int_{\Pi}p(\pi_k\mid\mathbf{y}_k,\pi_{k-1})p(\pi_{k-1},\mathbf{a}_{1:k-1},\mathbf{x}_{0:k-1})d\pi_{k-1}.\label{eq:joint_dist_init}
\end{align}
The Bayesian inverse filtering problem is defined as inferring the CR's posterior belief \(\pi_k\), given \(\mathbf{x}_{0:k}\) and \(\mathbf{a}_{1:k}\), expressed as \cite{KM_Patta2021CRmodel}:
\begin{equation}\label{eq:BIF_prob}
    \rho_k(\pi) = p(\pi_k \mid \mathbf{a}_{1:k}, \mathbf{x}_{0:k}).
\end{equation}
Analogous to the forward Bayesian filtering recursion in \eqref{eq:BayesFilter}, the posterior update in Bayesian inverse filtering in \eqref{eq:BIF_prob} can be recursively written, based on \eqref{eq:dynamics_mot} through \eqref{eq:joint_dist_init}, as \cite{KM_Patta2021CRmodel}:
\begin{equation}\label{eq:BIF_recurs}
   \hspace{-0.7em} \rho_{k+1}(\pi) =
    \frac{
    G_{\pi,\mathbf{a}_{k+1}}
    \int_{\Pi}
    B_{\mathbf{x}_{k+1},\mathbf{y}_{\pi_k,\pi}}
    \rho_k(\pi_k)\, d\pi_k
    }{
    \int_{\Pi} G_{\pi,\mathbf{a}_{k+1}}
    \int_{\Pi}
    B_{\mathbf{x}_{k+1},\mathbf{y}_{\pi_k,\pi}}
    \rho_k(\pi_k)\, d\pi_k \, d\pi
    },
\end{equation}
where \(\mathbf{y}_{\pi_k,\pi}\) is the observation \(\mathbf{y}\) such that \(\pi = T(\pi_k,\mathbf{y})\), with \(T\) defined in \eqref{eq:BayesFilter}. If \(P, B, \mathcal{A}\), and \(G\) are known to the learner, the recursion in \eqref{eq:BIF_recurs} can be implemented using the classical Bayesian inverse filtering techniques, namely, Inverse Hidden Markov Model Filter (IHMMF), Inverse Kalman Filter (IKF), Inverse Extended Kalman Filter (IEKF), Inverse Unscented Kalman Filter (IUKF), or Inverse Particle Filter (IPF), depending on the problem dynamics \cite{krishnamurthy2019calibrate,KM_Patta2021CRmodel,IPF2025IITD}.

\subsection{Bayesian Inverse Particle Filtering Framework}\label{sec:BIPF_framework}
Given the CR--learner interaction dynamics in \eqref{eq:dynamics_mot}–\eqref{eq:joint_dist_init} and the learner’s assumptions in Table~\ref{tab:assumptions}, the optimal inverse filtering recursion in \eqref{eq:BIF_recurs} generally lacks a finite-dimensional sufficient statistics.
Below we describe the inverse particle filter (IPF) algorithm proposed in \cite{krishnamurthy2019calibrate,IPF2025IITD}. 
The IPF empirically evaluates the recursive Bayesian posterior in \eqref{eq:BIF_recurs} using a set of N particles \(\{(\pi_{0:k}^{(i)},\mathbf{y}_{1:k}^{(i)})\}_{i=1}^N\) along with their corresponding importance weights \(\{w_k^{(i)}(\pi_{0:k}^{(i)},\mathbf{y}_{1:k}^{(i)})\}_{i=1}^N\), randomly evolving in time according to the dynamics of the CR--learner interaction \eqref{eq:dynamics_mot} through \eqref{eq:BayesFilter}.

The posterior distribution of the CR's belief \(\pi_{0:k}\) and observations \(\mathbf{y}_{1:k}\) given \(\mathcal{D}_{1:k} = \{\mathbf{x}_{0:k},\mathbf{a}_{1:k}\}\) can be expressed based on \eqref{eq:joint_dist_init} as \cite{krishnamurthy2019calibrate}:
\begin{align}
       p(\pi_{0:k},&\mathbf{y}_{1:k}\mid \mathbf{x}_{0:k}, \mathbf{a}_{1:k}) \propto G_{\pi_k, \mathbf{a}_k}p(\pi_k\mid \pi_{k-1},\mathbf{y}_k)\times\notag
       \\&p(\mathbf{y}_k\mid \mathbf{x}_k) p(\pi_{0:k-1},\mathbf{y}_{1:k-1}\mid \mathbf{x}_{0:k-1},\mathbf{a}_{1:k-1}).\label{eq:posterior_latent_init}
\end{align}Since drawing samples directly from \eqref{eq:posterior_latent_init} is intractable, IPF samples \(\pi_k^{(i)}\) and \(\mathbf{y}_k^{(i)}\) are instead drawn from a proposal distribution (importance density)\cite{krishnamurthy2019calibrate}:
\begin{align}
    Q(\pi_{0:k},\mathbf{y}_{1:k}&\mid \mathbf{x}_{0:k}, \mathbf{a}_{1:k}) =\notag\\& Q(\pi_{0:k-1},\mathbf{y}_{1:k-1}\mid \mathbf{x}_{0:k-1},\mathbf{a}_{1:k-1})\label{eq:prop_dist_init}\\&
    Q(\pi_k,\mathbf{y}_k\mid \pi_{0:k-1},\mathbf{y}_{1:k-1},\mathbf{x}_{0:k}, \mathbf{a}_{1:k}).\notag
\end{align} Given the original--\eqref{eq:posterior_latent_init}, and, the proposal--\eqref{eq:prop_dist_init} distributions, the importance weight in IPF can be defined as\cite{krishnamurthy2019calibrate}:
\begin{equation}\label{eq:imp_weight_init}
    w_k^{(i)}(\pi_{0:k}^{(i)},\mathbf{y}_{1:k}^{(i)}) = \frac{p(\pi_{0:k}^{(i)},\mathbf{y}_{1:k}^{(i)}\mid \mathbf{x}_{0:k}, \mathbf{a}_{1:k})}{Q(\pi_{0:k}^{(i)},\mathbf{y}_{1:k}^{(i)}\mid \mathbf{x}_{0:k}, \mathbf{a}_{1:k})}.
\end{equation} In this particular problem, we use the following optimal importance density \cite{krishnamurthy2019calibrate,IPF2025IITD} that minimizes the variance of importance weights in \eqref{eq:imp_weight_init}:
\begin{align}
    Q^*(\pi_k^{(i)},&\mathbf{y}_k^{(i)}\mid \pi_{0:k-1}^{(i)},\mathbf{y}_{1:k-1}^{(i)},\mathbf{x}_{0:k},\mathbf{a}_{1:k})\notag\\&=
    p(\pi_k^{(i)},\mathbf{y}_k^{(i)}\mid \pi_{0:k-1}^{(i)},\mathbf{y}_{1:k-1}^{(i)},\mathbf{x}_{0:k},\mathbf{a}_{1:k}).\label{eq:optim_impo_density}
\end{align} Using \eqref{eq:posterior_latent_init}, \eqref{eq:prop_dist_init} and \eqref{eq:optim_impo_density} the importance weight computation in \eqref{eq:imp_weight_init} reduces to the following recursive form:
\begin{align}
    w_k^{(i)}(\pi_{0:k}^{(i)},\mathbf{y}_{1:k}^{(i)})
    &\propto 
    w_{k-1}^{(i)}(\pi_{0:k-1}^{(i)},\mathbf{y}_{1:k-1}^{(i)})\notag\\
    &\qquad\times 
    G_{\pi_k^{(i)},\mathbf{a}_k}\, P_{\mathbf{x}_{k-1},\mathbf{x}_k}.\label{eq:importance_weight_recursion}
\end{align}Bayesian inverse filtering algorithm, presented in Algorithm~\ref{BIF_existing} assume that \(P, B\), \(G\) and \(T\) in \eqref{eq:dynamics_mot} through \eqref{eq:dynamics_act} are perfectly known to the learner. 

\begin{algorithm}[t]
{\fontsize{9pt}{9pt}\selectfont
\caption{Bayesian Inverse Particle Filter \cite{krishnamurthy2019calibrate,IPF2025IITD}}
\label{BIF_existing}
\begin{algorithmic}[1]

\REQUIRE Observed data $\mathbf{x}_{0:k}$, $\mathbf{a}_{1:k}$; 
number of particles $N$; 
known models $P, B, T, G$

\STATE Initialize particles $\{\pi_0^{(i)}\}_{i=1}^N \sim \pi_0$, 
set $w_0^{(i)} = \frac{1}{N}$

\FOR{$j = 1$ to $k$}

    \FOR{$i = 1$ to $N$}

        \STATE Sample $\mathbf{y}_j^{(i)} \sim B_{\mathbf{x}_j,\mathbf{y}}$
        
        \STATE Propagate belief:
        \[
        \pi_j^{(i)} = T(\pi_{j-1}^{(i)}, \mathbf{y}_j^{(i)})
        \]

        \STATE Update importance weight using \eqref{eq:importance_weight_recursion}:
        \[
        w_j^{(i)} \propto 
        w_{j-1}^{(i)} 
        G_{\pi_j^{(i)},\mathbf{a}_j} 
        P_{\mathbf{x}_{j-1},\mathbf{x}_j}
        \]

    \ENDFOR

    \STATE Normalize weights:
    \[
    w_j^{(i)} = \frac{w_j^{(i)}}{\sum_{\ell=1}^{N} w_j^{(\ell)}}
    \]

    \STATE Resample $\{\pi_{0:j}^{(i)}\}_{i=1}^N$
    according to $\{w_j^{(i)}\}_{i=1}^N$

\ENDFOR

\STATE \textbf{Output:} Weighted particle approximation
\[
p(\pi_{0:k},\mathbf{y}_{1:k}\mid \mathbf{x}_{0:k},\mathbf{a}_{1:k})
\approx 
\sum_{i=1}^N w_k^{(i)}
\delta(\pi_{0:k}-\pi_{0:k}^{(i)})
\]

\end{algorithmic}
}
\end{algorithm}

However, as indicated in Table~\ref{tab:assumptions}, \(\mathcal{A}\) and \(G\) are not known to the learner in our setting, and therefore any Bayesian inverse filtering algorithm must jointly infer both the belief-dependent radar policy (action likelihood) \(G_{\pi,\mathbf{a}} = p(\mathbf{a}\mid \pi)\), and the CR's posterior belief \(\pi\). \emph{Since parametric modeling of the action likelihood is not straightforward when the action space \(\mathcal{A}\) or its cardinality \(\lvert \mathcal{A} \rvert\) is unknown, in this paper, we adopt a Bayesian nonparametric approach.}
An important aspect in Bayesian non-parametric inference of \(G_{\pi,\mathbf{a}}\) is the choice of prior for \(G_{\pi,\mathbf{a}}\). In our context, since the CR selects its actions from a set of finitely many discrete actions, a Dirichlet distributed prior is a possible candidate, as is commonly used in Bayesian machine learning. However, since \(\lvert \mathcal{A} \rvert\) is unknown, the action likelihood prior must be modeled using a Dirichlet Process (DP), allowing the number of actions to grow to \(\infty\) as time progresses. Sec.~\ref{sec:DP_basics} provide a brief background of DP.

\subsection{Dirichlet Process}\label{sec:DP_basics}
The DP provides a natural and mathematically principled option for the prior over \(G_{\pi,\mathbf{a}}\).
\subsubsection{Definition of Dirichlet Process}\label{sec:DP_definition} Let \((\mathcal{A}, \mathcal{U})\) be a measurable space that represents possible action choices of the CR. A random probability measure \(G \sim \text{DP}(\alpha, H)\) with concentration parameter \(\alpha > 0\) and a base measure \(H\) is called a Dirichlet process if, for every measurable partition \((\mathcal{U}_1, \cdots, \mathcal{U}_m)\) of \(\mathcal{A}\) \cite{ferguson1973bayesianDP},
\begin{equation}\label{eq:DP_def}
    \hspace{-1.1em}\left( G(\mathcal{U}_1), \cdot\cdot, G(\mathcal{U}_m) \right)
    \sim \text{Dir}\left( \alpha H(\mathcal{U}_1), \cdot\cdot\cdot, \alpha H(\mathcal{U}_m) \right),
\end{equation}
where \(\text{Dir}(\cdot)\) is the Dirichlet distribution of order \(m\) with parameters \(\boldsymbol{\alpha} = (\alpha H(\mathcal{U}_1), \cdots, \alpha H(\mathcal{U}_m))\). Its realizations \((\nu_{1}, \cdots, \nu_{m})\) are \(m\)-dimensional vectors belonging to the \((m-1)\)-dimensional probability simplex, with probability density function\cite{mosimann1962DirichPDF}:
\begin{equation}\label{eq:DirchletPDF}\hspace{-0.9em}
    f(\nu_{1},\cdot\cdot,\nu_{m};\boldsymbol{\alpha}) = \frac{\Gamma(\sum_{i=1}^{m}\alpha H(\mathcal{U}_i))}{\prod_{i = 1}^{m}\Gamma(\alpha H(\mathcal{U}_i))} \prod_{i = 1}^{m}\nu_{i}^{\alpha H(\mathcal{U}_i)-1}.
\end{equation} In \eqref{eq:DirchletPDF}, \(\nu_i = p(\mathbf{a}\in \mathcal{U}_i)\) and \(\Gamma(\cdot)\) is the standard Gamma function.
The definition of the DP in \eqref{eq:DP_def} implies that \(G\) is almost surely discrete, and therefore suitable for modeling categorical phenomena whose support can grow as more data are observed. 

Given \(G\), the generation of CR action in \eqref{eq:dynamics_act} can be modeled by the learner using the DP as \cite{ferguson1973bayesianDP}:
\begin{equation}
    \mathbf{a}_t \mid G {\sim} G; \; t = 1,\cdots,T.
\end{equation}
\emph{However, in our context, the radar's policy depends on its posterior belief of the target's kinematic states, and any model for radar's policy should depend on the belief \(\pi\), necessitating the use of a dependent Dirichlet Process (DDP) \cite{maceachern2001DDP_seminal}. The details of the proposed DDP prior modeling are provided in Sec.~\ref{sec:PropMeth}.}

\section{Proposed Methodology}\label{sec:PropMeth}
Initially we discuss, Bayesian non-parametric characterization of belief dependent policy using DDP in Sec.~\ref{sec:DDP_characterization}. Subsequently, in Sec.~\ref{sec:IPF_DDP_joint}, we incorporate the DDP characterization of the policy in Sec.~\ref{sec:DDP_characterization} within the inverse particle filtering framework in Sec.~\ref{sec:BIPF_framework} to jointly infer the belief and the belief dependent policy of the CR.

\subsection{Modeling of \(G_{\pi,\mathbf{a}}\) using Dependent Dirichlet Process}\label{sec:DDP_characterization}
\subsubsection{DDP formulation and generative model}
Since the CR's policy \(G\) depends on belief \(\pi \in \Pi\) - ideally, we need to introduce individual DP priors for each \(\pi \in \Pi\) - necessitating a DDP formulation - capable of characterizing random probability measure that changes depending on the covariate \(\pi\)\cite{maceachern2001DDP_seminal}. Extending the DP formulation in Sec.~\ref{sec:DP_basics}, the generative model for the DDP characterization can be represented for any \(\pi \in \Pi\) as follows:
\begin{equation}
    \begin{aligned}
    G_{\pi} &\sim \text{DP}(\alpha,H_{\pi}),\\
    \mathbf{a}\mid G_{\pi} &\sim G_{\pi}.\label{eq:DDP_gen_model}
\end{aligned}
\end{equation} \(H_{\pi}\) in \eqref{eq:DDP_gen_model} is a base probability measure (corresponding to belief \(\pi\)) over the action space \(\mathcal{A}\), and \(\text{DP}(\cdot)\) denotes the Dirichlet process as defined in Sec.~\ref{sec:DP_definition}. Since \(|\mathcal{A}|\) is not known to the learner apriori, we hierarchically model \(H_{\pi}\) in \eqref{eq:DDP_gen_model} as a DP over the unknown \(\mathcal{A}\) such that any partition \(\mathcal{U} \subset \mathcal{A}\) and its complement \(\mathcal{U}^{C}\) will be Dirichlet distributed as defined in \eqref{eq:DP_def}.

\subsubsection{DDP Predictive Distribution}
A key property of the DP is its closed-form predictive distribution obtained by marginalizing out \(G\) \cite{maceachern2001DDP_seminal,Quintana2022DDP}.  
Assume we are given a set of \(T\) observations, comprising the CR's belief and corresponding actions, \(\{(\pi_k,\mathbf{a}_k)\}_{k=1}^{T}\). Let the radar actions be \(\mathbf{a}_{1:T}\) with distinct revealed categories \(\mathcal{A}_T=\{\mathbf{u}_1,\ldots,\mathbf{u}_{N_T}\}\), where \(N_T=|\mathcal{A}_T|\) and \(\mathcal{A}_T\cup\mathcal{A}_T^{C}=\mathcal{A}\).
For a new belief \(\pi_{T+1}\in\Pi\), the next action \(\mathbf{a}_{T+1}\) may either match an existing category (\(\mathbf{a}_{T+1}\in\mathcal{A}_T\)):
\begin{equation}
    \begin{aligned}\label{eq:DP_pred_dist_part1}
        p&(\mathbf{a}_{T+1} = \mathbf{u}_i \in \mathcal{A}_T \mid \mathbf{a}_{1:T}, \pi_{T+1})
        =\\& \frac{\alpha H_{\pi_{T+1}}(\mathbf{u}_i)+\sum_{k = 1}^{T}\boldsymbol{1}_{\{\mathbf{u}_k = \mathbf{u}_i,\pi_{T+1} = \pi_k\} }\delta_{\mathbf{u}_k}}{T + \alpha};
    \end{aligned}
\end{equation}or be a previously unseen one (\(\mathbf{a}_{T+1}\in\mathcal{A}_T^{C}\))
\begin{equation}
    \begin{aligned}\label{eq:DP_pred_dist_part2}
        p(\mathbf{a}_{T+1} = \mathbf{u} \in \mathcal{A}_T^C \mid \mathbf{a}_{1:T},\pi_{T+1})
        &= \frac{\alpha H_{\pi_{T+1}}(\mathbf{u})}{T + \alpha}.
    \end{aligned}
\end{equation}
The predictive distributions \eqref{eq:DP_pred_dist_part1} and \eqref{eq:DP_pred_dist_part2} naturally capture the possibility that the radar may take previously unrevealed action categories, with probability governed by the concentration parameter \(\alpha\). 

\subsubsection{DDP Posterior Update}After observing $\mathbf{a}_{1:T}$, the posterior over \(G_{\pi}\sim \text{DP}(\alpha, H_{\pi})\), \(\forall \pi\in \Pi\), remains a DP\cite{ferguson1973bayesianDP} given by:
\begin{align}
    G_{\pi}\mid \mathbf{a}_{1:T} &\sim\notag\\& \text{DP}\left(\alpha + T, \frac{\alpha H_{\pi}+ \sum_{t = 1}^{T}\boldsymbol{1}_{\{\pi_t = \pi\}}\delta_{\mathbf{a}_t}}{\alpha + T}\right).\label{eq:DP_posterior}
\end{align}This posterior is centered on a convex combination of the prior base measure and the empirical distribution revealed by the \(T\) observed actions.

However, in our context, the covariate—CR belief \((\pi_t\in\Pi)\) is not directly observable by the learner, and hence, \eqref{eq:DP_pred_dist_part1} through \eqref{eq:DP_posterior} cannot be used to approximate the likelihood \(G_{\pi, \mathbf{a}}\) in \eqref{eq:importance_weight_recursion}, since \(\pi\) and \(G\) must be jointly inferred. In Sec.~\ref{sec:IPF_DDP_joint}, we incorporate DDP inference of policy into the inverse particle filter framework to jointly infer \(p(G,\pi_{0:k}\mid\mathbf{x}_{0:k},\mathbf{a}_{1:k})\). Further, since the belief space \(\Pi\) is continuous, we use the standard kernel smoothing techniques to update the posterior from finite set of observations.

\subsection{IPFDDP: Joint inference of belief and policy}\label{sec:IPF_DDP_joint}

In this section, we describe the joint online inference of the radar's belief and the belief--dependent action policy by incorporating the DDP characterization of the radar's policy from Sec.~\ref{sec:DDP_characterization} into the Bayesian inverse particle filtering formulation of the radar belief estimate in Sec.~\ref{sec:BIPF_framework}. We jointly sample the latent CR belief \((\pi)\), the latent CR observation \((\mathbf{y})\), and the DP--distributed policy \(G\), given the model dynamics in \eqref{eq:dynamics_mot} and the Bayesian filter \eqref{eq:BayesFilter} used by the CR. A straightforward extension of the importance density in \eqref{eq:optim_impo_density} is used to jointly sample the particles:
\begin{align}
    Q^*&(\pi_k^{(i)}, \mathbf{y}_k^{(i)}, G_{0:k}^{(i)}
    \mid \pi_{0:k-1}^{(i)}, \mathbf{y}_{1:k-1}^{(i)}, G_{0:k-1}^{(i)}, \mathcal{D}_{1:k}) \notag\\[2pt]
    &= p(\pi_k^{(i)}, \mathbf{y}_k^{(i)}, G_{0:k}^{(i)}
    \mid \pi_{0:k-1}^{(i)}, \mathbf{y}_{1:k-1}^{(i)}, G_{0:k-1}^{(i)}, \mathcal{D}_{1:k}) \notag\\[2pt]
    &= p(G_{0:k}^{(i)} \mid \pi_{0:k}^{(i)}, G_{0:k-1}^{(i)}, \mathcal{D}_{1:k})\,
       \delta\!\bigl(\pi_k^{(i)} - T(\pi_{k-1}^{(i)}, \mathbf{y}_k^{(i)})\bigr) \notag\\
    &\phantom{=}\;\times\, p(\mathbf{y}_k^{(i)} \mid \mathbf{x}_k),
    \qquad i = 1, \ldots, N,
    \label{eq:optim_impo_density_joint}
\end{align}
where \(\mathcal{D}_{1:k} = \{\mathbf{x}_{0:k}, \mathbf{a}_{1:k}\}\) denotes the trajectory of the observed state--action pairs. The functions \(T(\pi_{k-1}^{(i)}, \mathbf{y}_k^{(i)})\) and \(p(\mathbf{y}_k^{(i)} \mid \mathbf{x}_k)\) follow directly from the known dynamics in \eqref{eq:dynamics_mot} through \eqref{eq:dynamics_act}. The weights of each particle are updated as in (13), wherein the $G_{\pi,k}$ in (13) is replaced by $G_{0:k}^{(i)}$
For each particle \(i\), we initialize \(G_0^{(i)}\) as
\begin{equation}
    G_0^{(i)} \sim \text{DP}(\alpha_0^{(i)}, H_0^{(i)}),\label{eq:DP_prior}
\end{equation}
with \(\alpha_0^{(i)} > 0\) as the concentration parameter and \(H_0^{(i)}\) as the base measure defined on the entire action space \(\mathcal{A}\). Although the action space \(\mathcal{A}\) or \(|\mathcal{A}|\) is not explicitly known, after each observation (\(k = 0, 1, \ldots\)), we may partition \(\mathcal{A}\) into the observed actions \(\mathcal{A}_k\) and their unobserved complement \(\mathcal{A}_k^C\), and thereby specify a probability measure via the DP as in \eqref{eq:DP_def}. The probability of observing an action not previously observed is captured, up to a proportionality constant, through the concentration parameter and an initial non-informative base measure. As the learner receives additional observations, the posterior belief over \(G\) is updated non-parametrically, and the posterior term in \eqref{eq:optim_impo_density_joint} is again characterized by a DP. Conditioned on the particle history \(\pi_{0:k}^{(i)}\) and the observed actions \(\mathbf{a}_{1:k}\), the posterior over \(G_0^{(i)}\) satisfies
\begin{equation}
    \hspace{-1em}G_{0:k}^{(i)} \mid \pi_{0:k}^{(i)}, \mathbf{a}_{1:k}, G_{0}^{(i)} 
    \sim \text{DP}(\alpha_{0:k}^{(i)}(\pi), H_{0:k}^{(i)}(\pi)),
    \label{eq:DP_update_short_hand}
\end{equation}
where \(\alpha_{0:k}^{(i)}\) is the updated concentration parameter and \(H_{0:k}^{(i)}\) is the updated base measure according to \eqref{eq:DP_posterior}. 

{\sl \textbf{Smoothing the estimated radar policy}: }The radar policy estimated using the particle filter in \eqref{eq:DP_update_short_hand}, as a weighted combination of $G_{0:k}^{(i)}$, is inherently sample-based and therefore exhibits discontinuities over the belief space, making it unsuitable for practical ECM applications. Hence, in this paper, we apply, standard kernel smoothing techniques to obtain a regularized nonparametric posterior update.
Under the assumption that the policy varies smoothly with respect to the belief, we use a squared exponential kernel to capture the dependence of the policy on \(\pi\):
\begin{equation}
    K(\pi, \pi')
    = \sigma_f^2
      \exp\!\left(
        -\frac{\|\pi - \pi'\|^2}{2\ell^2}
      \right),\label{eq:SE_Kernel}
\end{equation}
where \(\sigma_f^2 > 0\) denotes the \emph{signal variance}, controlling the amplitude of the function, and \(\ell > 0\) is the \emph{length--scale} (bandwidth), governing how rapidly the function varies with respect to \(\pi\). Thus, the updated concentration parameter \(\alpha_{0:k}^{(i)}\) in \eqref{eq:DP_update_short_hand} is
\begin{equation}
    \alpha_{0:k}^{(i)}(\pi) 
    = \alpha_0^{(i)} 
      + \sum_{j=1}^{k} K(\pi, \pi_j^{(i)}),
    \qquad \forall\, i,\;\forall\, \pi \in \Pi,\label{eq:alphaUpdate}
\end{equation}
and \(\forall\, \pi \in \Pi\), the updated base measure  \(H_{0:k}^{(i)}\) in \eqref{eq:DP_update_short_hand} is
\begin{equation}
    H_{0:k}^{(i)}(\pi) 
    = \frac{
        \alpha_0^{(i)} H_0^{(i)}
        + \sum_{j=1}^{k} \delta_{\mathbf{a}_j} K(\pi, \pi_j^{(i)})
      }{
        \alpha_0^{(i)} 
        + \sum_{j=1}^{k} K(\pi, \pi_j^{(i)})
      };\forall\, i.\label{eq:HUpdate}
\end{equation}

Further, the importance weight, \(w_{0:k}^{(i)}(\pi_{0:k}^{(i)},\mathbf{y}_{1:k}^{(i)},G_{0:k}^{(i)})\) can be updated as a straight forward extension of \eqref{eq:importance_weight_recursion}, with \(G_{\pi_k^{(i)},\mathbf{a}_k} \sim G_{0:k-1}^{(i)}\). The likelihood of observing a previously observed action, \(\mathbf{a}_k\in \mathcal{A}_{k-1}\) or an unobserved action \(\mathbf{a}_k\in \mathcal{A}_{k-1}^C\) can be approximated by extending \eqref{eq:DP_pred_dist_part1} and \eqref{eq:DP_pred_dist_part2}, respectively, as follows:
    \begin{align}
        p(&\mathbf{a}_{k} = \mathbf{u} \in \mathcal{A}_{k-1} \mid \mathbf{a}_{1:k}, \pi_{k}^{(i)})\label{eq:DP_pred_dist_part1_joint}
        \\&= \frac{\sum_{j = 1}^{k-1}\boldsymbol{1}_{\{\mathbf{u} = \mathbf{a}_j\}}K(\pi_j^{(i)},\pi_k^{(i)}) + \alpha_0^{(i)} H_{0}^{(i)}(\mathbf{u})}{\alpha_0^{(i)}+ \sum_{j = 1}^{k-1}K(\pi_j^{(i)},\pi_k^{(i)})};\;\forall i,\notag
    \end{align}
and, 
\begin{equation}
    \begin{aligned}\label{eq:DP_pred_dist_part2_joint}
        p(\mathbf{a}_{k} = &\mathbf{u} \in \mathcal{A}_k^C \mid \mathbf{a}_{1:k},\pi_{k}^{(i)})
        \\&= \frac{\alpha_0^{(i)} H_{0}^{(i)}(\mathbf{u})}{\alpha_0^{(i)}+ \sum_{j = 1}^{k-1}K(\pi_j^{(i)},\pi_k^{(i)})};\;\forall i.
    \end{aligned}
\end{equation}
However, in practice, for large values of \(k\), \(\mathcal{A}_k \rightarrow \mathcal{A}\) and \(\mathcal{A}_k^C \rightarrow \emptyset\), and hence \( p(\mathbf{a}_{k} = \mathbf{u} \in \mathcal{A}_k^C \mid \mathbf{a}_{1:k},\pi_{k}^{(i)})\rightarrow 0\).

The proposed approach, IPFDDP, for joint inference of the CR's belief and policy is given in Algorithm~\ref{alg:IPFDDP}. The details of the DDP posterior update, corresponding to steps 10 through 12 of Algorithm~\ref{alg:IPFDDP}, are presented in Algorithm~\ref{alg:Posterior_update}. Furthermore, IPFDDP employs the non-parametric likelihood approximation described in Algorithm~\ref{alg:Likelihood_approx} to update the weights of the individual particles. At any time step \(k\), IPFDDP provides the particles \(\{\pi_{0:k}^{(i)}, G_{0:k}^{(i)}\}_{i=1}^{N}\) along with their corresponding importance weights \(\{w_{0:k}^{(i)}\}_{i=1}^{N}\). The combined measure is then expressed as a convex combination of the individual particles, as presented in Algorithm~\ref{alg:Global_Policy_approx}, the global policy approximation algorithm.
For any  measurable sets \(\psi \in \Pi\) and \(\mathcal{U} \in \mathcal{A}\), the belief dependent 
policy is characterized by:
\begin{equation}
    G_{0:k}(\psi, \mathcal{U}) = \sum_{i=1}^{N} w_{0:k}^{(i)} 
    G_{0:k}^{(i)}(\psi, \mathcal{U}),\label{eq:GlobalDDPUpdate}
\end{equation}
with
\begin{equation}
\begin{aligned}
    \mathbb{E}[G_{0:k}(\psi, \mathcal{U})]
    &= \mathbb{E}\left[\sum_{i=1}^{N} w_{0:k}^{(i)}
    G_{0:k}^{(i)}(\psi, \mathcal{U})\right] \\
    &= \sum_{i=1}^{N} w_{0:k}^{(i)} 
    H_{0:k}^{(i)}(\psi, \mathcal{U}),
\end{aligned}\label{eq:ExpectedPolicy}
\end{equation}
and,
\begin{align}
\mathrm{Var}\!\left[ G_{0:k}(\psi,\mathcal{U}) \right]
&=
\mathrm{Var}\!\left[
\sum_{i=1}^{N} w_{0:k}^{(i)} \,
G_{0:k}^{(i)}(\psi,\mathcal{U})
\right] \nonumber \\
&=
\sum_{i=1}^{N} \left(w_{0:k}^{(i)}\right)^2
\mathrm{Var}\!\left[
G_{0:k}^{(i)}(\psi,\mathcal{U})
\right],
\label{eq:GlobalVariance}
\end{align}
where independence across particles is assumed. For each particle-level Dirichlet process,
\begin{equation}
\mathrm{Var}\!\left[
G_{0:k}^{(i)}(\psi,\mathcal{U})
\right]
=
\frac{
H_{0:k}^{(i)}(\psi,\mathcal{U})\,
\bigl(1 - H_{0:k}^{(i)}(\psi,\mathcal{U})\bigr)
}{
\alpha_{0:k}^{(i)}(\psi) + 1
}.
\label{eq:ParticleDPVariance}
\end{equation}

\begin{algorithm}[t]
{\fontsize{9pt}{9pt}\selectfont
\caption{Sequential DDP-Based Inverse Particle Filtering Algorithm (IPFDDP)}
\label{alg:IPFDDP}
\begin{algorithmic}[1]

\REQUIRE Initial sets: $\mathcal{A}_0 = \emptyset$, $\alpha_0 = 1$, $H_0$, $\pi_0 = (\mathbf{x}_0, \Sigma_0)$, $R$,
\FORALL{\(i\)}
\STATE weights $w_0^{(i)} = \frac{1}{N}$;

\FOR{$k = 1$ to $K$}

    \STATE Observe $\mathbf{x}_k$ and $\mathbf{a}_k$%

    \FOR{$i = 1$ to $N$}

        \STATE $\mathbf{y}_k^{(i)} \sim p(\mathbf{y} \mid \mathbf{x}_k)$
        \STATE $\pi_k^{(i)} \gets T(\pi_{k-1}^{(i)},\, \mathbf{y}_k^{(i)})$

        \STATE Update: $\alpha_{0:k}^{(i)}$ using \eqref{eq:alphaUpdate}
        \STATE Update: $H_{0:k}^{(i)}$ using \eqref{eq:HUpdate}

        \STATE 
        \(
        G_{0:k}^{(i)} \mid \pi_{0:k}^{(i)}, \mathbf{a}_{1:k}
        = \mathrm{DP}\!(\alpha_{0:k}^{(i)},\, H_{0:k}^{(i)})
        \): use Alg.~\ref{alg:Posterior_update}
        \STATE Compute: \(G_{\pi_k^{(i)},\, \mathbf{a}_k} = G_{0:k-1}^{(i)}\!(\pi_k^{(i)},\, \mathbf{a}_k )\): use Alg.~\ref{alg:Likelihood_approx}
        \STATE Update: \(
        w_k^{(i)} \propto
        w_{k-1}^{(i)} \;
        G_{0:k-1}^{(i)}\!(\pi_k^{(i)},\, \mathbf{a}_k )
        \)
        \STATE Update: \(\mathcal{A}_k \gets \mathcal{A}_{k-1} \cup \mathbf{a}_k\)

    \ENDFOR

\ENDFOR
\ENDFOR
\STATE \textbf{Outputs:}\\ 
\textbf{Particle set:} $\{\pi_{0:K}^{(i)}, G_{0:K}^{(i)}, w_K^{(i)}\}_{i=1}^N$, \(\mathcal{A}_K\), \(\mathbf{a}_{1:K}\)
\end{algorithmic}
}
\end{algorithm}
\begin{algorithm}[t]
{\fontsize{9pt}{9pt}\selectfont
\caption{Non parametric posterior update of DDP}
\label{alg:Posterior_update}
\begin{algorithmic}[1]

\REQUIRE \(\mathbf{x}_{0:k}\), \(\mathbf{a}_{1:k}\), \(\alpha_0^{(i)}\), \(H_0^{(i)}\) and \(\pi_{0:k}^{(i)}\)
\FORALL{$\pi\in \Pi$ and $i$}
\FOR{$j = 1$ to $k$}
    \STATE Compute \( K(\pi,\pi_j^{(i)})\) using \eqref{eq:SE_Kernel}
    \STATE Compute \(\alpha_{0:j}^{(i)}(\pi)\) using \eqref{eq:alphaUpdate}
    \STATE Compute \(H_{0:j}^{(i)}(\pi)\) using \eqref{eq:HUpdate}
\ENDFOR
\STATE \(G_{0:k}^{(i)}(\pi) = \text{DP}(\alpha_{0:k}^{(i)}(\pi),H_{0:k}^{(i)}(\pi))\)
\ENDFOR

\STATE \textbf{Output:} \(G_{0:k}^{(i)}\)

\end{algorithmic}
}
\end{algorithm}
\begin{algorithm}[h]
{\fontsize{9pt}{9pt}\selectfont
\caption{Non parametric likelihood approximation}
\label{alg:Likelihood_approx}
\begin{algorithmic}[1]

\REQUIRE \(\pi_k^{(i)}, \mathbf{a}_k, \pi_{0:k-1}^{(i)}, \mathbf{a}_{1:k-1}, \mathcal{A}_{k-1}\)
\FORALL{\(i\)}
\FOR{\(j = 1\) to \(k-1\)}
    \STATE Compute \( K(\pi_{j}^{(i)},\pi_k^{(i)})\) using \eqref{eq:SE_Kernel} and store
\ENDFOR
    \IF{\(\mathbf{a}_k\in \mathcal{A}_{k-1}\)}
        \STATE Compute \(p(\mathbf{a}_k\mid \mathbf{a}_{1:k-1}, \pi_k^{(i)})\) using \eqref{eq:DP_pred_dist_part1_joint}
    \ELSE
        \STATE Compute \(p(\mathbf{a}_k\mid \mathbf{a}_{1:k-1}, \pi_k^{(i)})\) using \eqref{eq:DP_pred_dist_part2_joint}
    \ENDIF
    \ENDFOR
\STATE \textbf{Output:} \(G_{0:k-1}^{(i)}(\pi_k^{(i)},\mathbf{a}_k)\gets\)\(p(\mathbf{a}_k\mid \mathbf{a}_{1:k-1}, \pi_k^{(i)}) \;\forall i\)

\end{algorithmic}
}
\end{algorithm}
\begin{algorithm}[h]
{\fontsize{9pt}{9pt}\selectfont
\caption{Global policy approximation from IPFDDP particles}
\label{alg:Global_Policy_approx}
\begin{algorithmic}[1]

\REQUIRE \textbf{Particle set:} $\{\pi_{0:K}^{(i)}, G_{0:K}^{(i)}, w_K^{(i)}\}_{i=1}^N$, \(\mathcal{A}_K\), \(\mathbf{a}_{1:K}\), \(\pi \in \Pi\), \(u \in \mathcal{A}\)
\STATE \(G_{0:K}(\pi,u)\gets 0\)
\FOR{\(i = 1\) to \(N\)}
    \STATE Compute \(G_{0:K}^{(i)}(\pi, u)\) using Alg.~\ref{alg:Likelihood_approx}
    \STATE \(G_{0:K}(\pi,u)\gets w_K^{(i)}\times G_{0:K}^{(i)}(\pi, u)\)
\ENDFOR
\STATE \textbf{Output: }\(G_{0:K}(\pi,u)\)

\end{algorithmic}
}
\end{algorithm}

\subsection{Active sensing strategy in IPFDDP}\label{sec:ActiveSensing}
In the electronic warfare context, the inverse learner interacts with an adversarial radar to learn the radar's perception and perception-action policy, and it is essential to minimize the interaction and perform learning using the smallest possible number of observations (interactions), since an increase in interaction time makes the inverse learner more susceptible to ECCM attacks.
Accordingly, in this section, we extend the IPFDDP framework to an \emph{active sensing} setting, in which the learner actively selects its actions (e.g., acceleration maneuvers) to probe a desired belief region \(\pi_{k+1}\), observes the resulting action \(\mathbf{a}_{k+1}\), and updates the posterior over policies accordingly, thereby enabling sample--efficient learning. Active sensing, as referred to here, determines the belief that should be explored next, and the inverse learner is assumed to know an approximate mapping from the desired belief to an inverse learning action. Further details on how this is achieved are provided in Sec.~\ref{sec:NumActiveSensing}.

\begin{proposition}[Entropy--Concentration Based Active Probing]
\label{prop:ActiveProbing}
Let \(G_{0:k}\) denote the global belief--dependent policy posterior defined in
\eqref{eq:GlobalDDPUpdate}.  
For a given belief \(\pi \in \Pi\), define the predictive mean action
distribution
\begin{equation}\label{eq:pred_mean_acq}
    \bar{p}_{0:k}(\mathbf{a}\mid \pi)
=
\mathbb{E}[G_{0:k}(\pi,\mathbf{a})]
=
\sum_{i=1}^{N} w_{0:k}^{(i)}
H_{0:k}^{(i)}(\pi,\mathbf{a}).
\end{equation}
Define the acquisition function
\begin{equation}\label{eq:acq_fn}
    {\gamma}_{0:k}(\pi)
=
\frac{
\mathcal{H}\!\left(
\bar{p}_{0:k}(\cdot \mid \pi)
\right)
}{
\sum_{i=1}^{N} w_{0:k}^{(i)}\,
\alpha_{0:k}^{(i)}(\pi)
},
\end{equation}
where \(\mathcal{H}(\cdot)\) denotes Shannon entropy:
\begin{equation}
\mathcal{H}\!\left(\bar{p}_{0:k}(\cdot \mid \pi)\right)
=
- \sum_{\mathbf{a} \in \mathcal{A}}
\bar{p}_{0:k}(\mathbf{a} \mid \pi)
\log \bar{p}_{0:k}(\mathbf{a} \mid \pi),
\label{eq:ShannonEntropyAcquisition}
\end{equation}
and
\(\alpha_{0:k}^{(i)}(\pi)\) is the particle-level concentration parameter in
\eqref{eq:alphaUpdate}.
Selecting the next belief probe as
\begin{equation}\label{eq:optimal_pi_to_probe}
\pi_{k+1} = \arg\max_{\pi \in \Pi} {\gamma}_{0:k}(\pi)
\end{equation}
prioritizes beliefs regions exhibiting high predictive ambiguity in the global policy
while discounting regions of high posterior confidence.
\end{proposition}

{\sl \textbf{Rationale}: }When no observations are available, the belief--dependent policy is characterized by the uninformative prior in~\eqref{eq:DP_prior}, which corresponds to a maximum--entropy distribution.
As learning progresses and observations are accumulated within a given belief region, the posterior concentrates around the actual value, leading to a reduction in entropy. Moreover, due to the use of kernel smoothing in the IPFDDP posterior, beliefs lying near policy transition regions will have higher entropy, as multiple local policies contribute non-negligible mass masking clear action preferences. \emph{So, the effective concentration parameter in the denominator of the acquisition function, \eqref{eq:acq_fn}, promotes exploration of sparsely probed belief regions, and the entropy term ensures that the transition regions are sufficiently probed.} 

\begin{algorithm}[t]
{\fontsize{9pt}{9pt}\selectfont
\caption{Active sensing strategy for IPFDDP}
\label{alg:ActiveProbeSelection}
\begin{algorithmic}[1]
\REQUIRE Particle weights \(\{w_{0:k}^{(i)}\}_{i=1}^{N}\),
particle-level posteriors
\(\{H_{0:k}^{(i)}, \alpha_{0:k}^{(i)}\}_{i=1}^{N}\),
candidate belief set \(\Pi_{\mathrm{cand}}\)
\FORALL{\(\pi \in \Pi_{\mathrm{cand}}\)}
    \STATE Compute predictive mean using \eqref{eq:pred_mean_acq}
    \STATE Compute \({\gamma}_{0:k}(\pi)\) using \eqref{eq:acq_fn} and \eqref{eq:ShannonEntropyAcquisition}
\ENDFOR
\STATE Select next probe:
\[
\pi_{k+1}
=
\arg\max_{\pi \in \Pi_{\mathrm{cand}}}
{\gamma}_{0:k}(\pi)
\]
\RETURN \(\pi_{k+1}\), Optimal inverse learning action
\end{algorithmic}}
\end{algorithm}

\section{Numerical Results}\label{sec:Results}

We consider a setting involving a target on the learner’s side moving at nearly constant velocity, being tracked by an adversarial CR. In order to track the kinematic state
\(\mathbf{x}_k = [r_k,\ \dot{r}_k]^\top\) at time step \(k\), where \(r_k\) and \(\dot{r}_k\) denote the range and range rate of the target, respectively, the CR employs a Kalman filter based on the following motion model:
\begin{equation}
    \mathbf{x}_{k+1}
    = A_k\, \mathbf{x}_k + \mathbf{w}_k,
    \quad 
    \mathbf{w}_k \sim \mathcal{N}(\mathbf{0},\, Q_k),
    \label{eq:Num_MotionModel}
\end{equation}
where
\begin{equation}
    A_k =
    \begin{bmatrix}
        1 & \Delta t_k \\
        0 & 1
    \end{bmatrix},
    \quad
    Q_k
    = \sigma_a^2
    \begin{bmatrix}
        \frac{\Delta t_k^4}{4} & \frac{\Delta t_k^3}{2} \\
        \frac{\Delta t_k^3}{2} & \Delta t_k^2
    \end{bmatrix}.
    \label{eq:Num_AQDef}
\end{equation}

Here, \(A_k\) denotes the state transition matrix and \(Q_k\) denotes the nominal process noise covariance corresponding to a constant-velocity motion model driven by white noise with variance \(\sigma_a^2\), with \(\Delta t_k\) representing the fixed radar revisit interval between observations \(k-1\) and \(k\).

The CR gets a noisy observation of \(\mathbf{x}_k\) given by the linear Gaussian observation model:
\begin{equation}
    \mathbf{y}_k = H\, \mathbf{x}_k + \mathbf{v}_k,
    \qquad 
    \mathbf{v}_k \sim \mathcal{N}(\mathbf{0},\, R),\label{eq:Num_ObsModel}
\end{equation}
where
\begin{equation}
    H = 
    \begin{bmatrix}
        1 & 0 \\[2pt]
        0 & 1
    \end{bmatrix},
    \qquad
    R = 
    \begin{bmatrix}
        \sigma_r^2 & 0 \\[2pt]
        0 & \sigma_{\dot r}^2
    \end{bmatrix}.\label{eq:Num_HRDEf}
\end{equation} The matrices \(H\) and \(R\) in \eqref{eq:Num_HRDEf} represent the observation matrix and the observation noise covariance matrix, respectively. In our context, \(R\) is a function of the CR's sensing parameters and waveform choice as in \cite{sira2022advances}, and can be reconstructed by the learner given the waveform choice of the CR. We assume that the motion and observation models in \eqref{eq:Num_MotionModel}--\eqref{eq:Num_HRDEf} are known to both the CR and our side.

Based on the linear Gaussian motion and observation model assumption, the CR uses Kalman filter, \eqref{eq:Num_KFPred1} through \eqref{eq:Num_KFUpd3}, to estimate \(\pi_k = p(\mathbf{x}_k\mid \mathbf{y}_{1:k})\):
\begin{align}
    \hat{\mathbf{x}}_{k|k-1} &= A_k\,\hat{\mathbf{x}}_{k-1|k-1},\label{eq:Num_KFPred1} \\
    \Sigma_{k|k-1} &= A_k\,\Sigma_{k-1|k-1}\,A^\top_k + Q_k,\label{eq:Num_KFPred2}
\end{align}
\begin{align}
    S_k &= H\,\Sigma_{k|k-1}\,H^\top + R,\label{eq:Num_KFUpd0}\\
    K_k &= \Sigma_{k|k-1}\,H^\top \big( S_k \big)^{-1},\label{eq:Num_KFUpd1} \\
    \hat{\mathbf{x}}_{k|k} &= 
        \hat{\mathbf{x}}_{k|k-1}
        + K_k\big( \mathbf{y}_k - H\,\hat{\mathbf{x}}_{k|k-1}\big),\label{eq:Num_KFUpd2} \\
    \Sigma_{k|k} &= 
        (I - K_k H)\,\Sigma_{k|k-1}.\label{eq:Num_KFUpd3}
\end{align}
The CR's posterior estimate of \(\mathbf{x}_k\), obtained using Kalman Filter tracker is in turn Gaussian and can be parameterized using the mean state estimate, \(\hat{\mathbf{x}}_k\), and the covariance matrix \(\Sigma_k\). We define the predicted (prior) belief \(\pi_{k\mid k-1}\) and the updated (posterior) belief \(\pi_{k\mid k}\) as in \eqref{eq:Num_Belieparam}:
\begin{equation}
    \pi_{k|k-1} \triangleq 
    \Big(
        \hat{\mathbf{x}}_{k|k-1},\;
        \Sigma_{k|k-1}
    \Big),
    \;
    \pi_k \triangleq 
    \Big(
        \hat{\mathbf{x}}_{k|k},\;
        \Sigma_{k|k}
    \Big).\label{eq:Num_Belieparam}
\end{equation}
\begin{figure}[t]
    \includegraphics[width=\linewidth]{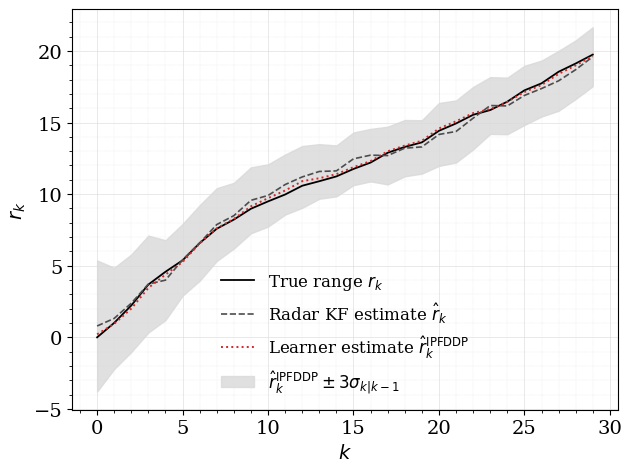}
    \caption{The range (\(r_k\)) estimate obtained using CR Kalman filter tracker and the IPFDDP estimate of CR's belief. Since the generative models used by the CR are assumed to be known to the inverse learner, the uncertainty of the learner regarding the CR's mean estimate of the target state is attributed to the IPFDDP particles \(\mathbf{y}_k^{(i)}\) representing the latent CR observations.}
    \label{fig:TrackingUncertaintyIPF_1}
\end{figure}
\begin{figure}[t]
    \includegraphics[width=\linewidth]{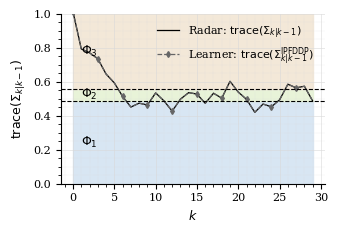}
    \caption{Trace of predictive covariance of CR's Kalman filter tracker and its estimate obtained by the learner using IPFDDP. The figure also shows the disjoint partitions of the CR's belief space as in \eqref{eq:Num_BeliefParts}. The radar takes action according to \eqref{eq:Num_PolicyMapping}: chooses LFM, PFM and HFM respectively for low, medium and high uncertainties respectively. \(\tau_1\) and \(\tau_2\) in \eqref{eq:Num_BeliefParts} are chosen in our context as \(0.49\) and \(0.56\) respectively.}
    \label{fig:TrackingUncertaintyIPF_2}
\end{figure}
Let \(\mathcal{A} = \{u_1, u_2, u_3\}\) be the action space (the set of all possible actions (i.e. waveform choices) CR can take). For example, \(u_1, u_2 \text{ and }u_3\) denote waveform chirps: Linear Frequency Modulation (LFM), Power Frequency Modulation (PFM) and Hyperbolic Frequency Modulation (HFM), respectively \cite{sira2022advances,sira2007dynamic,krishnamurthy2020CR_RP}.
In each time step, radar chooses the waveform based on the predicted belief \(\pi_{k\mid k-1} \in \Pi\) before making the observation \(\mathbf{y}_k\), where \(\hat{\mathbf{x}}_{k|k-1}\) given by \eqref{eq:Num_KFPred1} is the predicted state estimate and \(\Sigma_{k|k-1}\) given by \eqref{eq:Num_KFPred2} is the predicted covariance. CR quantifies the anticipated uncertainty in estimate as:
\begin{equation}\label{eq:uncertainty}
    {\sigma}_k = \text{trace}(\Sigma_{k|k-1}).
\end{equation} 
The CR partitions the belief space \(\Pi\) into three disjoint subsets, based on the predicted uncertainty as follows
\begin{equation}
    \begin{aligned}
        &\Phi_1 \triangleq \{\pi = (\hat{\mathbf{x}}, \Sigma)\in \Pi: \text{trace}(\Sigma)\le \tau_1\}\\
        &\Phi_2 \triangleq \{\pi = (\hat{\mathbf{x}}, \Sigma)\in \Pi: \tau_1 < \text{trace}(\Sigma)\le \tau_2\}\\
        &\Phi_3 \triangleq \{\pi = (\hat{\mathbf{x}}, \Sigma)\in \Pi: \text{trace}(\Sigma)> \tau_2\}.\\&
        \Phi_1 \cup \Phi_2 \cup \Phi_3 = \Pi,
    \qquad
    \Phi_i \cap \Phi_j = \varnothing,\quad i \neq j.
    \end{aligned}\label{eq:Num_BeliefParts}
\end{equation}

Let 
\(
G : \Pi \rightarrow \Delta(\{u_1,u_2,u_3\})
\)
denote the mapping from a belief \(\pi\) to the corresponding action–probability vector.  
The mapping
\begin{equation}
G_{\pi} =
\begin{cases}
(0.6,\,0.3,\,0.1), & \pi \in \Phi_1, \\[4pt]
(0.1,\,0.6,\,0.3), & \pi \in \Phi_2, \\[4pt]
(0.3,\,0.1,\,0.6), & \pi \in \Phi_3.
\end{cases}\label{eq:Num_PolicyMapping}
\end{equation} specifies how likely CR is to choose each action, given its predicted belief is \(\pi\). The conditional distribution is therefore:
\begin{equation}\label{eq:Num_Policy}
    a \mid \pi \sim G_{\pi}.
\end{equation}
The Kalman Filter tracking: \eqref{eq:Num_KFPred1} through \eqref{eq:Num_KFUpd3}, and the subsequent action selection strategy used in the simulation are illustrated in Fig.~\ref{fig:TrackingUncertaintyIPF_1} and Fig.~\ref{fig:TrackingUncertaintyIPF_2}, respectively. In the remainder of this section, we present the results of Radar belief and policy estimation using the proposed IPFDDP algorithm.
\begin{figure*}
\centering
    \includegraphics[width=\linewidth]{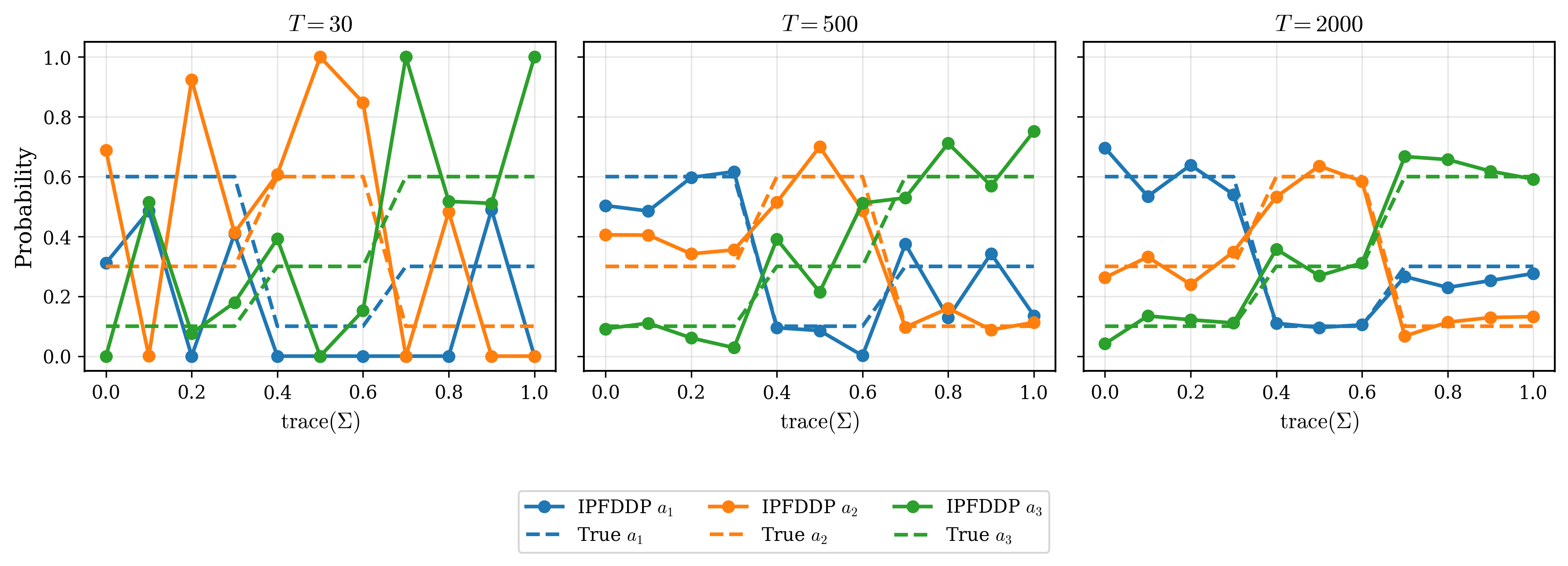}
    \caption{Mean of the approximated global policy learned via IPFDDP as a function of radar belief---covariance trace, for different data lengths $T \in \{30, 500, 2000\}$. Solid lines denote the mean estimated action probabilities, while dashed lines represent the true policy. As $T$ increases, the mean estimate converges to the true policy, demonstrating consistency of the proposed approach.}
    \label{fig:meanGlobalPolicyDDP}
\end{figure*}
\begin{figure*}
\centering
    \includegraphics[width=\linewidth]{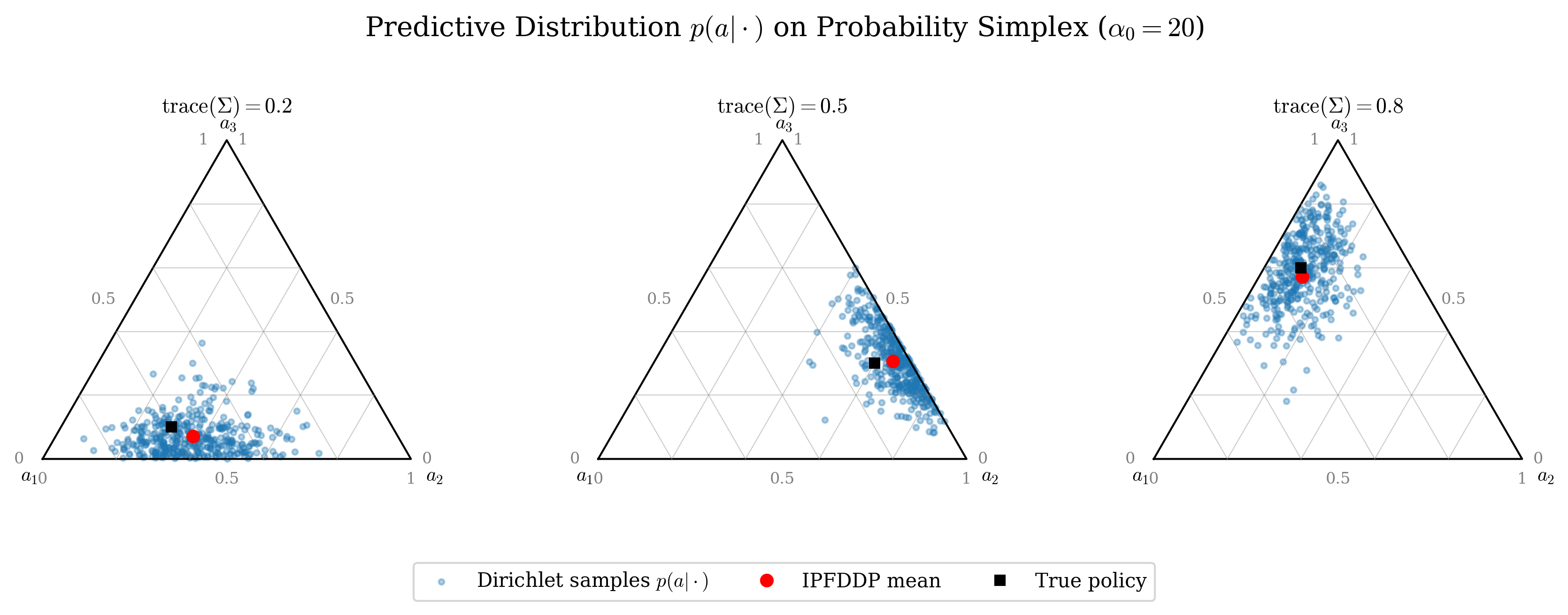}
    \caption{Samples on the probability simplex obtained from the Dirichlet process in \eqref{eq:GlobalDDPUpdate} for $\mathrm{trace}(\Sigma_{k\mid k-1})\in\{0.2,0.5,0.8\}$, corresponding to three belief-space partitions defined in \eqref{eq:Num_BeliefParts}. Blue points show Dirichlet samples, the red marker denotes the IPFDDP mean estimate, and the black marker indicates the true policy. For $T=2000$, the mean estimate closely matches the true policy. The spread of the samples depends on the initial concentration parameter $\alpha_0$ and the number of observations: larger $\alpha_0$ and increased data lead to higher confidence, consistent with the variance in \eqref{eq:GlobalVariance}.}
    \label{fig:samplesGlobalPolicyDDP}
\end{figure*}
\begin{figure*}
    \centering
    \includegraphics[width=\linewidth]{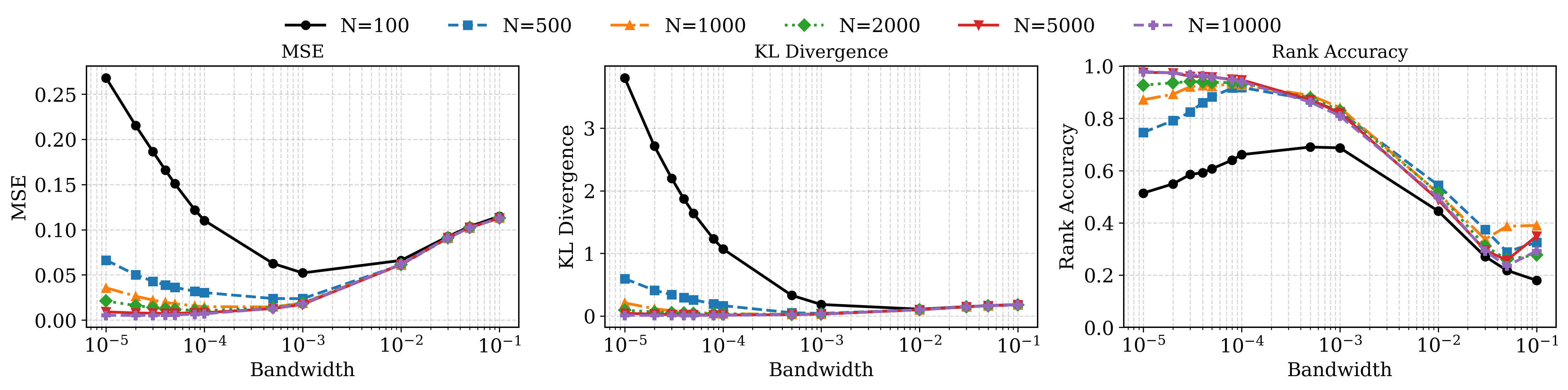}
    \caption{Model performance vs. bandwidth for varying training sample sizes (fixed test set = 200). Metrics: (a) MSE, (b) KL Divergence, (c) Rank Accuracy.}
    \label{fig:IPFPerformanceIPFDDP}
\end{figure*}

\subsection{Radar's belief estimated using IPFDDP}
We assume that the learner is equipped with a radar detector that can observe the CR's action, \(\mathbf{a}_k\), corresponding to the target's kinematic state, \(\mathbf{x}_k\). We further use the data set \(\mathcal{D}_{1:T} = \{\mathbf{x}_{0:T}, \mathbf{a}_{1:T}\}\), together with knowledge of the CR's generative model in \eqref{eq:Num_MotionModel}--\eqref{eq:Num_KFUpd3}, to jointly estimate the CR's belief of the target's kinematic state and the CR's belief-dependent action policy using the proposed IPFDDP algorithm described in Algorithm~\ref{alg:IPFDDP}. The results of inverse filtering, corresponding to the estimation of the CR's perception, are shown in Fig.~\ref{fig:TrackingUncertaintyIPF_1} through Fig.~\ref{fig:TrackingUncertaintyIPF_2}. Since the generative models used by the CR are assumed to be known to the inverse learner, the uncertainty of the learner regarding the CR's mean estimate of the target state in Fig.~\ref{fig:TrackingUncertaintyIPF_1} is attributed to the particles \(\mathbf{y}_k^{(i)}\) representing the latent CR observations. Given that \(A\), \(Q\), \(H\), \(R\), and \(T(\cdot)\) are perfectly known to the learner, the CR's tracking covariance can be accurately estimated using the IPFDDP algorithm. 

 In the remainder of this section, we discuss the effectiveness of IPFDDP in learning the belief dependent policy of the adversarial CR.

 \subsection{Belief dependent CR policy estimated using IPFDDP}\label{sec:IPFDDP_policy_Estimation_result}
 We used IPFDDP (Algorithm~\ref{alg:IPFDDP}) and the global policy approximation procedure in Algorithm~\ref{alg:Global_Policy_approx} to estimate the belief dependent CR policy given by equations \eqref{eq:Num_BeliefParts} through \eqref{eq:Num_Policy}, for various number of observations, \(T\), for initial concentration parameter, \(\alpha_0 = 20\), and for kernel bandwidth of \(l = 10^{-4}\). Algorithm~\ref{alg:Global_Policy_approx} approximates the CR policy using the DDP in \eqref{eq:GlobalDDPUpdate}, and the mean and samples respectively of the approximated global policy are visualized in Figures \ref{fig:meanGlobalPolicyDDP} and \ref{fig:samplesGlobalPolicyDDP}. Fig.~\ref{fig:meanGlobalPolicyDDP} demonstrates that the mean estimate, \eqref{eq:ExpectedPolicy}, converges to the true policy, \eqref{eq:Num_PolicyMapping} as \(T\) increases.

 Fig.~\ref{fig:samplesGlobalPolicyDDP} illustrates samples on the probability simplex drawn from the DDP in \eqref{eq:GlobalDDPUpdate} for covariance trace values of \(0.2\), \(0.5\), and \(0.8\), corresponding to three different partitions of the belief space as defined in \eqref{eq:Num_BeliefParts}. For \(T = 2000\), the mean estimate (red marker) lies very close to the true value (black marker), and the spread of the samples depends on the number of samples as well as the initial concentration parameter \(\alpha_0\). The confidence of the learner's policy estimate can be controlled by increasing \(\alpha_0\), which results in Dirichlet samples that are more tightly clustered around the mean value. Furthermore, for a fixed \(\alpha_0\), the learner becomes increasingly confident in its estimated policy as it observes a larger number of samples, as implied by the variance of the estimate in \eqref{eq:GlobalVariance}.
\subsection{Effect of bandwidth and number of observations on IPFDDP based CR policy predictions - random probing}
We used \(T = 100, 500, 1000, 2000, 5000, \text{and}\ 10000\) distinct observations to learn the CR's policy and to predict the policy corresponding to the subsequent \(K = 200\) observations using Algorithms~\ref{alg:IPFDDP} and~\ref{alg:Global_Policy_approx}.  
We compared the expected value of the learned CR policy from \eqref{eq:ExpectedPolicy} with the actual policy defined by \eqref{eq:Num_BeliefParts}--\eqref{eq:Num_PolicyMapping}, using the metrics:
\begin{inparaenum}[(i)]
    \item Mean Squared Error (MSE),
    \item KL Divergence, and
    \item accuracy in capturing the preferences towards action (denoted as RankAcc),
\end{inparaenum}defined in equations \eqref{eq:MSE} through \eqref{eq:RankAcc} respectively.
\begin{table}[t]
\centering
\caption{Performance Metrics Across Bandwidth Values}
\label{tab:bandwidth_results}
\begin{tabular}{c c c c}
\hline
\textbf{Bandwidth} & \textbf{MSE} & \textbf{KL} & \textbf{RankAcc} \\
\hline
$10^{-6}$  & 0.060269 & 0.726982 & 0.810 \\
$10^{-5}$  & 0.028568 & 0.230592 & \textbf{0.910} \\
$10^{-4}$  & \textbf{0.0048} & \textbf{0.0354} & \textbf{0.950} \\
$10^{-3}$  & 0.018648 &\textbf{ 0.033537} & 0.830 \\
$10^{-2}$  & 0.059849 & 0.097742 & 0.530 \\
$10^{-1}$  & 0.112931 & 0.180199 & 0.650 \\
$10^{0}$   & 0.125722 & 0.199276 & 0.055 \\
$10^{1}$   & 0.127164 & 0.201425 & 0.000 \\
\hline
\end{tabular}
\end{table}

Let $\{\pi_t\}_{t=1}^{K}$ denote a set of test belief points.
For each $\pi_t$, let the posterior mean CR policy inferred via IPFDDP be
\begin{equation}
\bar{p}_t
\triangleq
\mathbb{E}\!\left[p(\cdot \mid \pi_t)\mid\mathcal{D}\right]
\in \Delta^2,
\qquad
p_t^{\star}
\triangleq
p_{\text{true}}(\cdot \mid \pi_t),
\end{equation}
where $\Delta^2$ is the 3-dimensional probability simplex.

\noindent\textbf{1) Mean Squared Error (MSE):}
The MSE between the posterior mean policy and the true policy is defined as
\begin{equation}\label{eq:MSE}
\mathrm{MSE}
=
\frac{1}{K}
\sum_{t=1}^{K}
\left\|
\bar{p}_t - p_t^{\star}
\right\|_2^2 .
\end{equation}

\noindent\textbf{2) KL Divergence:}
Interpreting both $\bar{p}_t$ and $p_t^{\star}$ as categorical distributions over the action space, the average KL divergence is given by
\begin{equation}\label{eq:KL}
\mathrm{KL}
=
\frac{1}{K}
\sum_{t=1}^{K}
\sum_{i=1}^{3}
p_{t,i}^{\star}
\log
\frac{p_{t,i}^{\star}}{\bar{p}_{t,i}},
\end{equation}
where $p_{t,i}^{\star}$ and $\bar{p}_{t,i}$ denote the probabilities assigned to action $u_i$ by $p_t^{\star}$ and $\bar{p}_t$, respectively.

\noindent\textbf{3) Rank Accuracy:}
Rank accuracy evaluates whether the posterior mean policy preserves the ordinal action preferences.
Let $\mathrm{rank}(\cdot)$ denote the ordering induced by sorting the action probabilities in descending order.
The average rank accuracy is defined as
\begin{equation}\label{eq:RankAcc}
\mathrm{RankAcc}
=
\frac{1}{T}
\sum_{t=1}^{T}
\mathbb{I}
\!\left[
\mathrm{rank}(\bar{p}_t)
=
\mathrm{rank}(p_t^{\star})
\right],
\end{equation}
where $\mathbb{I}[\cdot]$ is the indicator function.

The non-parametric likelihood–estimation steps employed in both IPFDDP (Algorithm~\ref{alg:IPFDDP}) and the Global policy approximation (Algorithm~\ref{alg:Global_Policy_approx}) use the squared exponential kernel given in \eqref{eq:SE_Kernel}. As with all non-parametric kernel-based techniques, the correctness of the prediction is sensitive to the choice of the bandwidth (length scale) hyperparameter.
Trends in variation of MSE, KL divergence, and rank accuracy—for different training samples and bandwidths, are depicted in Fig.~\ref{fig:IPFPerformanceIPFDDP}.
Table~\ref{tab:bandwidth_results} specifically presents how the metrics vary with kernel bandwidth when \(T = 2000\) samples are used for learning the policy. For bandwidths in the range \(10^{-4}\) to \(10^{-2}\), the IPFDDP-based policy–learning model achieves MSE and KL divergence on the order of \(10^{-2}\). However, a finite bandwidth results in interference/overlap in the predictions near the partition boundaries, preventing the rank accuracy from reaching \(100\%\).
Adaptive bandwidth selection approaches \cite{kochenderfer2022algorithms} can be used for optimal kernel bandwidth selection, and the discussion is beyond the scope of this paper.
\subsection{Comparison with state of the art ML/DL models in characterizing CR policy}
To the best of our knowledge, ours is the first work in the literature that jointly infers the CR's belief and the belief-dependent policy. Since the radar selects its action based on its belief of the target's state, and this belief is not known to the learner, a simple function mapping from the target's kinematic state \(\mathbf{x}_k\) to the corresponding action \(\mathbf{a}_k\) is insufficient to characterize the CR's controller strategy. Furthermore, the action space \(\mathcal{A}\) is not known apriori, which makes a straightforward use of parametric ML models intractable. However, assuming the action space is known apriori, the models in Table~\ref{tab:model_performance} all can serve as function approximators that learn a mapping from the target state \(\mathbf{x}\) (or sequences \(\mathbf{x}_{t-L+1:t}\) in the case of sequence models as LSTMs) to a probability distribution over the CR's discrete action set \(\mathcal{A} = \{u_1, u_2, u_3\}\). The details of all the models are available in the supplementary material.
\begin{table}[t]
\centering
\caption{Performance Comparison of Models}
\label{tab:model_performance}
\begin{tabular}{lccc}
\hline
\textbf{Model} & \textbf{MSE} & \textbf{KL Divergence} & \textbf{RankAcc} \\
\hline
Logistic Regression\cite{hastie2009elements_Supp_01_Static_ML}  & 0.0310 & 0.1450 & 0.622 \\
Random Forest\cite{breiman2001random_Supp_03_Random_Forest}        & 0.0741 & 0.4982 & 0.565 \\
Gradient Boosting\cite{friedman2001greedy_Supp_04_Grad_Boost}    & 0.0865 & 0.4289 & 0.598 \\
SVM (RBF)\cite{platt1999probabilistic_Supp_06_SVM_Proba}            & 0.0232 & 0.1142 & 0.687 \\
kNN\cite{hastie2009elements_Supp_01_Static_ML}                  & 0.0361 & 0.2913 & 0.603 \\
Gaussian Process\cite{williams2006gaussian_Supp_07_GPClassi}     & 0.0279 & 0.1332 & 0.648 \\
MLP Neural Net\cite{bishop2006pattern}       & 0.0297 & 0.1421 & 0.620 \\
LSTM\_lag2\cite{hochreiter1997long_Supp_08_LSTM}           & 0.0180 & 0.0874 & 0.802 \\
LSTM\_lag3\cite{hochreiter1997long_Supp_08_LSTM}           & 0.0208 & 0.0981 & 0.713 \\
LSTM\_lag4\cite{hochreiter1997long_Supp_08_LSTM}           & 0.0201 & 0.0944 & 0.788 \\
LSTM\_lag6\cite{hochreiter1997long_Supp_08_LSTM}           & 0.0209 & 0.0999 & 0.782 \\
LSTM\_lag10\cite{hochreiter1997long_Supp_08_LSTM}          & 0.0206 & 0.0992 & 0.733 \\
Transformer\_lag2\cite{vaswani2017attention_Supp_09_Transf_Atten}    & 0.0223 & 0.1054 & 0.703 \\
Transformer\_lag3\cite{vaswani2017attention_Supp_09_Transf_Atten}    & 0.0263 & 0.1260 & 0.680 \\
Transformer\_lag4\cite{vaswani2017attention_Supp_09_Transf_Atten}    & 0.0328 & 0.1567 & 0.582 \\
Transformer\_lag6\cite{vaswani2017attention_Supp_09_Transf_Atten}    & 0.0347 & 0.1664 & 0.513 \\
Transformer\_lag10\cite{vaswani2017attention_Supp_09_Transf_Atten}   & 0.0457 & 0.2190 & 0.372 \\
Attention\_lag2\cite{vaswani2017attention_Supp_09_Transf_Atten}      & 0.0304 & 0.1440 & 0.622 \\
Attention\_lag3\cite{vaswani2017attention_Supp_09_Transf_Atten}      & 0.0321 & 0.1544 & 0.620 \\
Attention\_lag4\cite{vaswani2017attention_Supp_09_Transf_Atten}      & 0.0367 & 0.1742 & 0.472 \\
Attention\_lag6\cite{vaswani2017attention_Supp_09_Transf_Atten}      & 0.0384 & 0.1831 & 0.452 \\
Attention\_lag10\cite{vaswani2017attention_Supp_09_Transf_Atten}     & 0.0407 & 0.1935 & 0.445 \\
\textbf{IPFDDP}      & \textbf{0.0048} & \textbf{0.0354} & \textbf{0.950} \\
\hline
\end{tabular}
\\[1em]{\footnotesize Note: The detailed model architectures, training objectives and policy prediction details of all above models are presented in supplementary material.}
\end{table}

During training, each model adjusts its parameters to minimize its respective loss, so that \(p_{\theta}(a \mid \mathbf{x})\) approximates the true underlying action probabilities \(\{p_1, p_2, p_3\}\) associated with the regions of the continuous feature space \(\mathcal{X}\). Once trained, the model outputs an estimate \(\hat{p}(a \mid \mathbf{x})\) for each action, learning how different \(\mathbf{x} \in \mathcal{X}\) correspond to distinct radar action likelihoods. Table~\ref{tab:model_performance} compares the performance of the IPFDDP-based policy–learning algorithm with state-of-the-art ML models, based on MSE, KL divergence, and rank accuracy, given by equations \eqref{eq:MSE} through \eqref{eq:RankAcc}. 
Among these methods, the LSTM variants show improved performance over classical ML models, with LSTM\_lag2 achieving the best results within the LSTM family. However, the proposed \textbf{IPFDDP} model significantly outperforms all baselines, achieving the lowest MSE(\(0.0048\)) and KL divergence(\(0.0354\)), and the highest rank accuracy (\(0.950\)), demonstrating its superior predictive capability and more accurate belief–action modeling.
 
The performance of IPFDDP can be further improved through optimal bandwidth selection and by explicitly incorporating active sensing strategies. \emph{Importantly, none of the other models in Table~\ref{tab:model_performance} can support active sensing, as they do not provide a principled quantification of uncertainty in the estimated belief or action distribution.}
\subsection{Effect of Active Sensing on IPFDDP Performance}\label{sec:NumActiveSensing}
\begin{figure}[t]
    \centering
    \includegraphics[width=\linewidth]{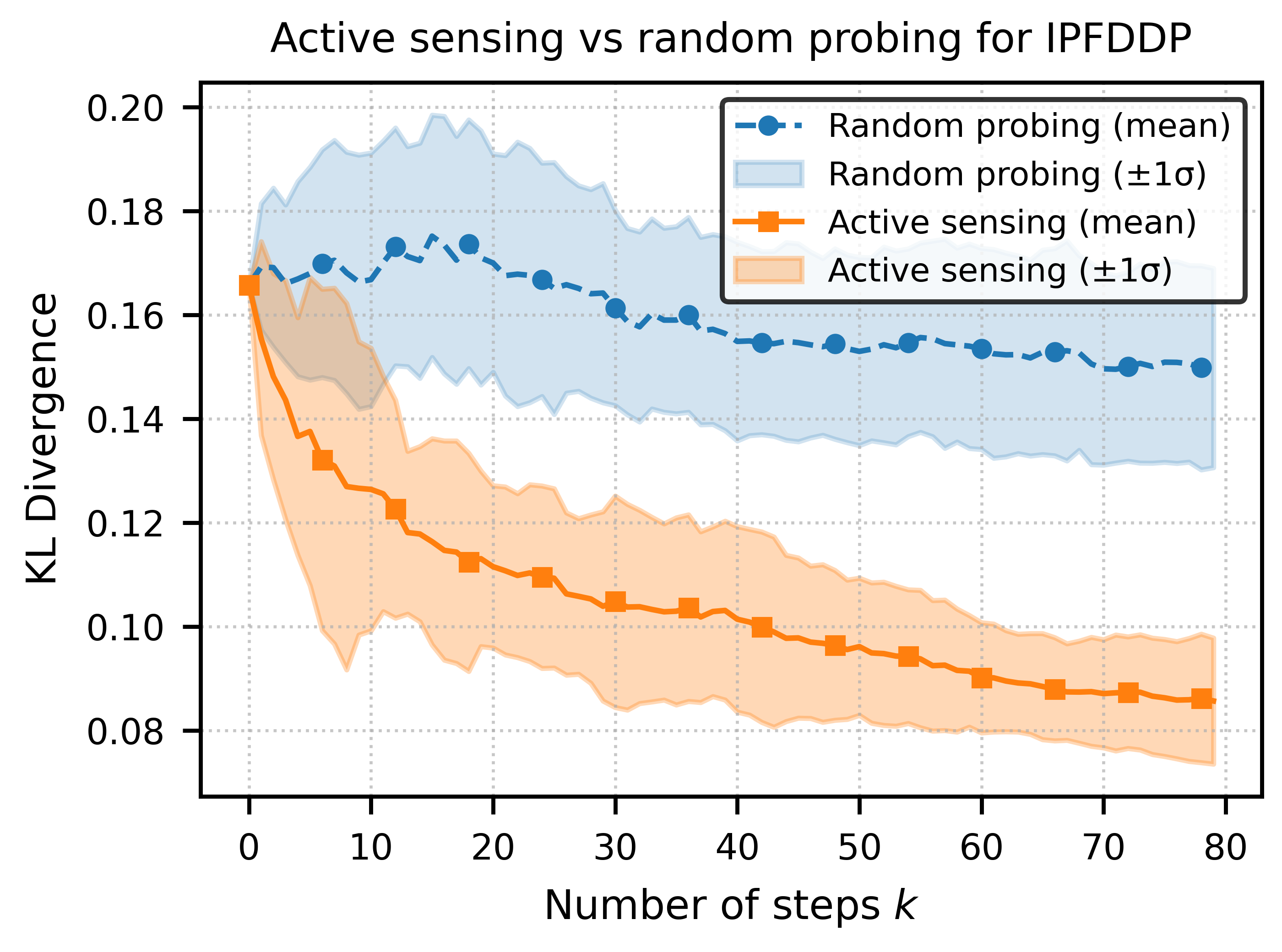}
    \caption{Reduction in KL divergence for IPFDDP under active sensing compared to random probing, showing faster convergence and lower uncertainty.}
    \label{fig:active_vs_random}
\end{figure}
In modern CR systems, optimal Bayesian trackers are often augmented with adaptive noise estimation or covariance-matching mechanisms that continuously adjust the effective process noise covariance associated with the assumed target dynamics model $P_{\mathbf{x}_{k-1},\mathbf{x}_k}$ \cite{krishnamurthy2020cognitive,mehra1970adaptiveKF}. 

In our setting, the CR employs the state-space model described in \eqref{eq:Num_MotionModel}--\eqref{eq:Num_KFUpd3} to track the target. Under the constant-velocity assumption, the predicted target state according to the CR's generative model is given by \eqref{eq:Num_KFPred1}. 
The CR estimates the target's acceleration maneuver through the innovation process
\begin{equation}\label{eq:innov}
    \nu_k = \mathbf{y}_k - H \mathbf{x}_{k\mid k-1}.
\end{equation}
An adaptive Kalman filtering routine \cite{mehra1970adaptiveKF,adapt_KF_GPS} matches the empirical innovation variance
\begin{equation}\label{eq:emp_innov_var}
    C_{\nu_K} = \mathbb{E}[\nu_k^\top \nu_k],
\end{equation}
with the theoretical innovation covariance $S_k$ defined in \eqref{eq:Num_KFUpd0}, by compensating for model mismatch through adjustment of the process noise covariance matrix $Q$, as in \cite{adapt_KF_GPS}:
\begin{equation}
    \hat{Q}_k = K_k \, C_{\nu_K} \, K_k^{\top},
\end{equation}
where $K_k$ is the Kalman gain matrix given in \eqref{eq:Num_KFUpd1}. 
Hence, the adapted predictive covariance becomes
\begin{equation}
    \hat{\Sigma}_{k|k-1} = A_k\,\Sigma_{k-1|k-1}\,A_k^\top + \hat{Q}_k,
\end{equation}
and the anticipated uncertainty in \eqref{eq:uncertainty} becomes
\begin{equation}
    \sigma_k = \mathrm{trace}\!\left(\hat{\Sigma}_{k\mid k-1}\right).
\end{equation}
The CR then takes actions according to \eqref{eq:Num_BeliefParts}--\eqref{eq:Num_PolicyMapping} based on the tracking uncertainty $\sigma_k$.
If the optimal belief to be probed according to \eqref{eq:optimal_pi_to_probe} is $\pi'$, the optimal acceleration maneuver $a_{k+1}$ for probing can be obtained by solving
\begin{equation}\label{eq:prob_x}
    \mathbf{x}_{k+1} = [r_{k+1},\dot{r}_{k+1}] 
    = \arg\min_{\mathbf{x}'} 
    \mathbb{E}\!\left[\pi' - \rho(\hat{\pi}_k, \mathbf{x}')\right],
\end{equation}
subject to the kinematic constraint
\begin{equation}\label{eq:prob_accel}
    r_{k+1} = r_k + \dot{r}_k \Delta t + \frac{a_{k+1} (\Delta t)^2}{2}.
\end{equation} The term, \(\rho(\pi_k,\mathbf{x}^{'})\), in \eqref{eq:prob_x} is given by the inverse filtering recursion in \eqref{eq:BIF_recurs} for \(\mathbf{x}_{k+1} = \mathbf{x}^{'}\).

When the true target acceleration deviates significantly from the constant-velocity assumption in \eqref{eq:Num_MotionModel}, persistent innovation bias arises. Under this paradigm, purposeful acceleration maneuvers by the target induce sustained mismatch relative to the CR's assumed motion model. Consequently, the adaptive covariance-matching mechanism modifies the effective process noise covariance used by the CR's tracker \cite{mehra1970adaptiveKF,adapt_KF_GPS}.

Fig.~\ref{fig:active_vs_random} illustrates the impact of the optimal probing strategy discussed in Proposition~\ref{prop:ActiveProbing} of Sec.~\ref{sec:ActiveSensing} on the performance of IPFDDP, measured in terms of the KL divergence (equation \eqref{eq:KL}) (mean and standard deviation over \(100\) trials) between the true policy and the IPFDDP estimate. The horizontal axis of Fig.~\ref{fig:active_vs_random} represents the number of time steps following \(200\) random CR--target interactions, for the experimental setting described in Sec.~\ref{sec:IPFDDP_policy_Estimation_result}. Active sensing consistently achieves a faster and more pronounced reduction in KL divergence (\(\approx 40\%\) faster) compared to random probing, indicating more informative data acquisition. In addition, the variance under active sensing is significantly lower, reflecting improved estimation stability across trials. These results confirm that uncertainty-aware probing substantially enhances belief refinement in IPFDDP.
\section{Conclusion}\label{sec:Conclusion}
This paper considered an adversarial inference problem involving an adversarial cognitive radar (CR) and a learner, in which the CR uses a Bayesian tracker to form a perception of the target kinematic states and takes possibly stochastic actions according to a perception-dependent policy, while the learner observes these actions corresponding to known target states and seeks to jointly infer the CR’s perception and policy. Motivated by the lack of existing learning frameworks that can simultaneously estimate both the perception and the perception-driven stochastic action policy of a cognitive agent, we proposed an online nonparametric Bayesian machine learning framework and developed the IPFDDP algorithm. IPFDDP jointly infers the belief (perception) and action-selection strategy of the CR by characterizing the belief-dependent policy using a Dependent Dirichlet Process and embedding kernel-based DDP inference within a Bayesian inverse particle filtering framework. Numerical simulations demonstrated that the proposed IPFDDP-based policy learning approach outperforms state-of-the-art machine learning models in terms of MSE, KL divergence between the estimated and true policy, and accuracy in identifying relative action preferences. 
The proposed Bayesian formulation inherently quantifies uncertainty in the learner’s estimates of perception and policy, enabling the design of active probing strategies in which the target dynamically adapts its acceleration maneuver to probe the CR in regions of maximum uncertainty. Simulation results show that incorporating active probing into IPFDDP reduces the KL divergence between the estimated and true policy, on average, $40\%$ faster than randomized probing, highlighting the benefits of uncertainty-aware active learning in adversarial cognitive radar settings.

\bibliographystyle{IEEEtran}
\bibliography{Cognitive_short}

\end{document}